\documentclass[10pt,conference]{IEEEtran}
\IEEEoverridecommandlockouts
\usepackage{amsmath,amssymb,amsfonts}
\usepackage{graphicx}
\usepackage{textcomp}
\usepackage{xcolor}
\usepackage{colortbl}
\usepackage{booktabs}
\usepackage{url}
\usepackage{framed}
\usepackage[hidelinks]{hyperref}
\hypersetup{
  pdftitle={The Hidden Costs of Forgetting: A Study of Trajectory Deletion in Software Engineering Agents},
  pdfauthor={Fernando Vallecillos-Ruiz, Islem Bouzenia, Michael Pradel, Leon Moonen},
  pdfkeywords={LLM agents, software engineering agents, trajectory reduction, inference cost}
}
\usepackage[abbreviate=true, dateabbrev=true, alldates=year, isbn=false, doi=true, urldate=comp, url=false, maxbibnames=10, backref=false, backend=biber, style=numeric-comp, language=american, giveninits=true, sortcites=true, sorting=none]{biblatex}

\usepackage{citation-cleaner}

\def\BibTeX{{\rm B\kern-.05em{\sc i\kern-.025em b}\kern-.08em T\kern-.1667em\lower.7ex\hbox{E}\kern-.125emX}}

\definecolor{shadecolor}{gray}{0.9}
\newenvironment{myframed}{%
  \MakeFramed {\advance\hsize-\width \FrameRestore}}%
{\endMakeFramed}

\newcommand{\bemph}[1]{\textbf{\emph{#1}}}

\definecolor{resultBetter}{RGB}{27,120,55}
\definecolor{resultWorse}{RGB}{190,45,45}
\newcommand{\resultcell}[2]{\cellcolor{#1}#2}

\begin{document}

\title{The Hidden Costs of Forgetting: A Study of Trajectory Deletion in Software Engineering Agents}

\author{
\IEEEauthorblockN{
Fernando Vallecillos-Ruiz\IEEEauthorrefmark{1},
Islem Bouzenia\IEEEauthorrefmark{2},
Michael Pradel\IEEEauthorrefmark{2},
Leon Moonen\IEEEauthorrefmark{1}
}

\IEEEauthorblockA{
\IEEEauthorrefmark{1}
\textit{Simula Research Laboratory}, Norway\\
fernando@simula.no, leon.moonen@computer.org
}

\IEEEauthorblockA{
\IEEEauthorrefmark{2}
\textit{CISPA Helmholtz Center for Information Security}, Germany\\
bouzenia.islem@pm.me, michael@binaervarianz.de
}
}

\maketitle

\noindent
\begin{abstract}
Software engineering agents resend their entire trajectory to an LLM at every turn, which motivates work on online trajectory reduction.
Prior work shows that even randomly deleting 75\% of the tokens in old turns may preserve the agent's effectiveness, but it remains unclear why this is the case, what the agent does to compensate for missing content, and whether trajectory reduction actually reduces costs.
This paper presents an empirical study of 42 trajectory-reduction configurations, focusing on their effects on effectiveness, agent behavior, and cost.
Our study systematically explores the effects of deleting content from old turns by varying the scope, granularity, amount, and timing of deletion across 9,400 runs with two state-of-the-art agents on the SWE-bench Verified dataset.
We find that deleting large amounts of trajectory content (e.g., up to 100\% of reasoning text) can preserve overall effectiveness.
Importantly, such deletion needs to be applied with a delay, as deleting content immediately after it is generated tends to significantly degrade effectiveness.
Agents often compensate for deleted content by re-fetching it, which can increase both the number of turns an agent takes to solve a task and the number of input and output tokens.
Whether trajectory reduction reduces LLM costs depends on the pricing model:
Under a flat pricing model, where all input tokens are billed equally, estimated costs fall in 26 of 42 evaluations across both agents.
Surprisingly, under a caching-aware pricing model, none of these 42 evaluations yields lower estimated costs, since reducing the trajectory breaks the cached prefix, whose tokens are billed at a lower rate.
Our findings have several important implications:
First, token counts alone are misleading; trajectory reduction techniques must instead consider overall costs under a realistic pricing model.
Second, studies of trajectories are important for understanding agent behavior, as focusing only on end-to-end effectiveness can hide important behavioral differences, such as excessive re-fetching of deleted content.
Finally, these results highlight the need for future work on memory management, that is, deciding which trajectory content to keep or delete and when to delete it, to reduce costs without sacrificing effectiveness.
\end{abstract}


\section{Introduction}

\noindent
Software engineering agents based on large language models (LLMs) increasingly automate software engineering tasks by inspecting repositories, modifying code, and validating patches~\cite{jimenez2024swebench}. Such agents retain prior reasoning, tool calls, and tool outputs in a trajectory that grows as they work. This growth increases the context processed by the model and can eventually exceed its context window~\cite{xiao2026:reducing,liu2026contexttool}, making trajectory management an important challenge.

Prior approaches reduce retained trajectories by discarding messages or tool outputs, selecting content to remove, or summarizing earlier interactions~\cite{xiao2026:reducing,liu2026contexttool,zhang2026:autocompact}. For example, AgentDiet reports that randomly deleting up to 75\% of eligible trajectory tokens can preserve much of an agent's effectiveness~\cite{xiao2026:reducing}. These studies demonstrate that agents can tolerate substantial trajectory reduction.

However, existing results do not show how the individual design choices in trajectory deletion affect agents. A deletion policy must decide what to remove, how much to remove, whether to remove whole turns or individual tokens, and whether to protect recent turns. Because trajectories contain diverse information, from intermediate reasoning to tool observations, these choices may affect agents differently. A systematic study that isolates deletion scope, granularity, amount, and timing can clarify how each choice affects task solving.

Trajectory reduction requires broader evaluation than final task success or token counts alone. An agent may complete a task by revisiting files, repeating commands, or retrieving deleted information, adding work even when the outcome is unchanged~\cite{liu2026:what,weinberger2026:token}. Token counts also do not establish cost savings: prompt caching discounts reusable input prefixes, and rewriting a trajectory can invalidate part of that prefix~\cite{openai2026promptcache,anthropic2026promptcache,xu2026:tokenpilot,weinberger2026:token}. Evaluations thus need to measure task effectiveness, subsequent agent behavior, and execution cost together.

This paper presents a systematic empirical study of trajectory deletion in software engineering agents. We vary scope, granularity, amount, and timing across 42 deletion configurations, evaluating each on two popular agents (Mini-SWE-agent~\cite{yang2024:sweagent} and Trae agent~\cite{gao2025:trae}). Half of the configurations delete with a two-turn delay and half without it, which gives 42 evaluations per timing across both agents combined. The study covers 100 SWE-bench Verified~\cite{jimenez2024swebench} tasks and 9,400 runs. We measure effectiveness as tasks solved, analyze how agents behave after content is deleted, and estimate execution costs under both flat and caching-aware pricing. More specifically, we address three research questions.

\bemph{RQ1: How do deletion scope, granularity, amount, and timing affect effectiveness and execution length?}
We find that the timing of trajectory reduction is critical.
With a two-turn delay, 37 of 42 evaluations do not statistically significantly reduce the number of tasks solved.
In contrast, immediate deletion significantly reduces the number of tasks solved in 19 of 42 evaluations.
The kind of deleted content also matters: Removing reasoning usually does not affect effectiveness, but deleting tool output is more harmful.
Finally, we observe that trajectory reduction tends to increase the number of turns an agent needs (in 21 of 42 and 34 of 42 evaluations with delayed and immediate deletion, respectively).

\bemph{RQ2: How does trajectory deletion affect agents' subsequent behavior, particularly the repeated retrieval of previously observed information?}
Agents often compensate for deleted content by re-fetching it.
For example, deleting tool outputs raises the re-fetch rate to as much as 7.5 times (with delayed deletion) and 10.6 times (with immediate deletion) compared to the baseline.

\bemph{RQ3: How does trajectory deletion affect estimated execution cost under flat and caching-aware pricing?}
The answer heavily depends on the pricing model.
Assuming a flat pricing model, where all input tokens are billed equally, 26 of 42 evaluations with the delay reduce estimated costs across both agents.
However, the results change under a caching-aware pricing model, where cached input prefixes are billed at a lower rate.
In this case, none of these 42 evaluations reduce estimated costs, and 26 evaluations are significantly more expensive than the uncompressed baseline.

Our findings have several important implications for future work.
First, trajectory reduction should be evaluated not only in terms of task success and token counts.
Instead, evaluations should also consider execution length, retrieval behavior, and estimated costs under realistic pricing models.
Second, it is important to study the behavior of agents after trajectory reduction, as they may compensate for deleted content by re-fetching it, which can increase execution length and costs.
Finally, we expect the detailed results across our 42 deletion configurations to inform future research on trajectory reduction techniques, including memory management strategies that decide what to keep or delete and when to delete it.

In summary, this work makes three contributions:
\begin{enumerate}
\item A systematic study of trajectory deletion in agents, quantifying how deletion scope, granularity, amount, and timing affect effectiveness, agent behavior, and execution costs.

\item A detailed analysis of trajectories resulting from trajectory deletion, showing that agents selectively re-fetch deleted tool outputs, preserved effectiveness can conceal longer executions, and fewer tokens can yield higher estimated costs under caching-aware pricing.

\item A dataset and replication package~\cite{replicationpackage} containing 9,400 trajectories produced by two agents on 100 SWE-bench Verified tasks using 42 trajectory reduction configurations, including the original content of deleted turns and analysis code for studying retrieval behavior and execution cost.

\end{enumerate}

\section{Background}
\label{sec:background}

\subsection{Agents and Their Trajectories}

\noindent
An LLM-based agent solves tasks, such as GitHub issues, through repeated model calls and repository interactions~\cite{yang2024:sweagent,wang2025:openhands,gao2025:trae}. At each turn, the model produces reasoning text and tool calls, such as commands to inspect files, modify code, or run tests. The agent executes these calls and records their outputs. A \emph{turn} comprises one model call, its reasoning, tool calls, and resulting outputs. The sequence of turns forms the agent's \emph{trajectory}~\cite{bouzenia2025:understanding}.
An agent retaining its complete trajectory includes previous interactions and newly acquired information in each request, so the prompt grows during execution. Tool outputs, including file contents, search results, and test logs, contribute to this growth. Since input-token processing is a major part of coding-agent costs~\cite{bai2026:how}, reducing the retained trajectory offers a way to limit context consumption.

\subsection{Trajectory Reduction}

\noindent
Trajectory-reduction approaches rewrite old turns before resending them and differ in what they remove and retain.
\emph{Masking} replaces old tool outputs with a placeholder~\cite{yang2024:sweagent,lindenbauer2025:complexity}.
\emph{Selection} uses a model to identify and remove irrelevant content~\cite{xiao2026:reducing,zeng2026:attncompress}.
\emph{Summarization} (compaction) replaces a span of turns with an LLM summary~\cite{smith2025openhands,anthropic2026contextediting}.
Recently, agents have been trained to decide when and what to compact~\cite{liu2026contexttool,zhang2026:autocompact}.
Our study builds on AgentDiet~\cite{xiao2026:reducing}, in which a reflection model removes content classified as useless, redundant, or expired from every turn longer than 500 tokens once it is two turns old.
AgentDiet's random baseline deletes 75\% of eligible tokens and solves almost as many tasks as its model-based reduction approach.
These results motivate studying agents' compensation for deletion and the resulting additional execution costs.

\begin{figure}[t]\centering
  \includegraphics[width=\columnwidth]{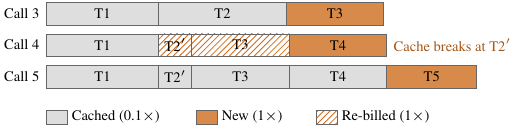}
  \caption{The longest prefix identical to the previous prompt is billed at the cached price. Reducing T2 at Call~4 rewrites it to T2$'$, i.e., T2$'$ and every later turn are re-billed. At Call~5, the prefix matches again and only T5 is new.}
  \label{fig:cache}
  \vspace*{-2ex}
\end{figure}

\subsection{Prefix Caching and Its Pricing}
\label{sec:bg-caching}

\noindent
A model computes key and value vectors for prompt tokens and stores them in a key--value (KV) cache.
When a prompt repeats the previous prompt and appends a turn, inference engines reuse cached states for the matching token prefix~\cite{zheng2024:sglang,vllm2026prefixcaching}.
Prefix reuse ends at the first token that differs because each cached state depends on the preceding tokens.
\mbox{OpenAI}, Anthropic, Google, and DeepSeek discount input tokens served from matching cached prefixes.
When the cache is available, an agent appending turns without rewriting history pays the cached price for the repeated prefix.
Rewriting an old turn changes the prefix and prevents reuse of the cached states from the first change onward (Figure~\ref{fig:cache}). Under the caching-aware pricing model used in this study, these tokens incur the full input price.
SWE-agent's documentation warns that eliding old observations breaks prompt caching~\cite{sweagent2026history}, and Anthropic's documentation states that clearing tool results or reasoning invalidates the cached prefix~\cite{anthropic2026contextediting}.
Section~\ref{sec:rq3} evaluates whether token savings offset the cost of processing tokens beyond the matching prefix.

\section{Study Design}
\label{sec:design}

\subsection{Agents and Tasks}
\label{sec:design-agents}

\noindent
We evaluate two agents with different tool interfaces.
The first is Trae agent \cite{gao2025:trae}, the agent used by AgentDiet~\cite{xiao2026:reducing}, which we include to support comparison with AgentDiet.
Trae agent features a structured file viewer, an editor, and a shell environment.
The second is Mini-SWE-agent \cite{yang2024:sweagent}, a setup restricted to a single shell tool, which provides a contrast to the specialized tools of Trae agent.
Both agents use the same model, DeepSeek-V4-Flash \cite{deepseek-ai2026:deepseekv4}.
We choose this open-weight model to support self-hosted inference, complete call logging, and execution of 9,400 runs (Section~\ref{sec:design-impl}).

The tasks are the fixed 100-instance subset of SWE-bench Verified \cite{jimenez2024swebench} used by AgentDiet~\cite{xiao2026:reducing}.
Each task corresponds to a real GitHub issue and has a hidden, human-validated test suite.
Using AgentDiet's subset~\cite{xiao2026:reducing} supports direct comparison and limits the experiment to 4,700 runs per agent.

\subsection{Deletion Configurations}
\label{sec:design-conditions}

\noindent
A turn yields three pieces of information: (i) reasoning text produced by the model, (ii) one or more tool calls suggested by the model, and (iii) the outputs of those tool calls.
The trajectory contains information from all completed turns and forms part of each subsequent model request.
We vary four dimensions of deletion: scope, granularity, amount, and timing.
Each \emph{configuration} assigns values to each dimension, and is evaluated on both agents. Deletion rewrites eligible turns before a model call, conforming to the configuration.

\textbf{Scope.} The scope is the part of a turn eligible for deletion.
We evaluate three scopes:
(i) reasoning text, (ii) tool output, or (iii) both, including the tool call. The third scope therefore covers the entire turn.
We do not consider the tool call itself as a scope because it is usually short.
Reasoning includes observations, plans for subsequent actions, and hypotheses about the cause of a bug.
Tool outputs include file contents, search results, and test logs.

Deletion occurs after a turn's tool calls execute, so the deletion configuration leaves commands unchanged before execution.

\textbf{Granularity.}
\emph{Turn-level} deletion (Turn in the tables and figures) drops the scoped content of a turn entirely, with probability $p$, and replaces it with a one-line note indicating deleted content.
\emph{Token-level} deletion (Token in the tables and figures) removes $100p\%$ of the scoped tokens of a turn, chosen at random, and keeps a note together with the surviving tokens.
At the same $p$, both granularities delete approximately the same expected percentage of scoped tokens.
Token-level deletion retains some tokens in each eligible turn, whereas turn-level deletion either removes or retains all scoped content.

After each turn, the procedure processes the newly eligible turn.
Turn-level deletion draws uniformly from $[0,1]$ and removes the scoped content if the draw is below $p$.
For token-level deletion, the procedure tokenizes the scoped content with a fixed tokenizer and removes $\lfloor p \cdot n \rfloor$ of its $n$ tokens, sampled uniformly without replacement.

\textbf{Amount.}
We express the amount $p \in [0,1]$ as a percentage of scoped tokens deleted, so $p=0.25$ corresponds to 25\% deletion.
We use 25\%, 50\%, 75\%, and 100\% for turn-level deletion, to cover the range from light to total deletion, and 25\%, 50\%, and 75\% for token-level deletion since at 100\% the two granularities coincide.
This gives $3 \times (4 + 3) = 21$ combinations of scope, granularity, and amount.

\begin{figure}[t]\centering
  \includegraphics[width=\columnwidth]{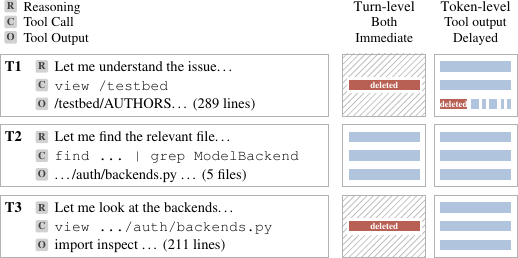}
  \caption{The first three turns (T1--T3, left) of a Trae agent trajectory and the prompts at the next call under two configurations (right). The first configuration deletes T1 and T3, while T2 remains intact. The second configuration modifies parts of T1.}
  \label{fig:deletion}
    \vspace*{-2ex}
\end{figure}

\textbf{Timing.}
A turn becomes eligible for rewriting after a fixed delay.
With \emph{delayed} timing, the delay is two turns, so the last two turns are always intact.
The two-turn delay follows AgentDiet~\cite{xiao2026:reducing}.
With \emph{immediate} timing, the delay is zero. This configuration tests the effect of removing the retention delay.
Immediate deletion rewrites a turn before the next model call, so the agent never observes the original content of a deleted output.
The procedure considers each turn once, when the turn becomes eligible.
Therefore, a turn that survives the turn-level draw stays intact, while a rewritten turn stays rewritten for the rest of the run.
Figure~\ref{fig:deletion} illustrates the two granularities and contrasts immediate and delayed timing on a real Trae agent trajectory.
The 21 combinations at each of two timings yield 42 configurations.
Evaluating every configuration on both agents gives 84 evaluations, with 42 evaluations per timing.

\subsection{Research Questions and Metrics}
\label{sec:design-rqs}
\label{sec:design-metrics}

\noindent
\textbf{RQ1. Effectiveness: How do deletion scope, granularity, amount, and timing affect effectiveness and execution length?}
To answer RQ1, we evaluate all 42 configurations (Section~\ref{sec:design-conditions}) on both agents and the same 100 tasks (Section~\ref{sec:design-agents}).
Furthermore, we add five uncompressed runs per agent.
Because inference is not deterministic, the repeated baselines characterize execution variability without deletion.
For each run, we track four metrics:
(1) a task is solved when the patch passes its associated test suite. A failing patch, no patch, or a run that ends with a context-length error counts as not solved. We report the number of tasks solved, or the solved rate when expressed as a percentage;
(2) run length is the number of turns, which we summarize by its median across tasks;
(3) a run is capped when it reaches the 200-turn budget. A capped run can still solve the task if it already holds a working patch while testing it;
(4) a run loops when it repeats an identical tool call at least ten times in a row.
We compare each configuration with the uncompressed runs (Section~\ref{sec:design-stats}).
We first examine scope, granularity, and amount under delayed deletion, following AgentDiet~\cite{xiao2026:reducing}, and then compare delayed and immediate deletion to assess the effect of timing.

\textbf{RQ2. Behavior: How does an agent compensate for content deleted from its trajectory?}
To answer RQ2, a rule-based labeler classifies each tool call as a read, edit, test, script, or other operation.
The labeler identifies file reads performed through shell commands or Python scripts.
We manually verify the labels for ten randomly sampled logs, five per agent. For each agent, the five sampled logs comprise one uncompressed run and runs from two delayed and two immediate configurations.
The labeler assigns 851 of the 855 labels correctly (99.5\% accuracy).
The replication package includes the classifier and the ten manually verified logs.
We then mark each read as a re-fetch or an initial read: a read is a re-fetch when at least 80\% of its distinct lines already appear in the output of an earlier read in the same run.
We exclude outputs under five lines to avoid accidental matching.
The re-fetch rate is the number of re-fetches divided by the number of reads, per run, reported as a ratio to the uncompressed runs (Section~\ref{sec:design-stats}).
Normalizing by the number of reads supports comparison across configurations with different run lengths.
The analysis also compares re-fetching of deleted and retained outputs and characterizes run behavior.
Furthermore, we measure on the uncompressed runs how far back an agent reaches for content.
For a re-fetch, we count the turns since its lines last appeared in a read.
For an edit, we count the turns since the agent last read the file.
The percentage of re-fetches and edits whose source lies within the last $k$ turns indicates what a deletion delay of $k$ turns would leave intact.

\textbf{RQ3. Cost: How does trajectory deletion affect estimated execution cost under flat and caching-aware pricing?}
To answer RQ3, we calculate token counts from the logs (Section~\ref{sec:design-impl}) and estimate each run's cost relative to the uncompressed baseline under two pricing models. Under \emph{flat pricing}, every prompt token costs the input price. Under \emph{caching-aware pricing}, only the longest prefix identical, token for token, to the previous call's prompt costs the cached price, and all remaining prompt tokens cost the input price (Section~\ref{sec:bg-caching}). Both models bill completion tokens at the output price. An uncompressed trajectory preserves the prefix, whereas a rewrite re-bills tokens from the first change onward. Turn-level deletion breaks the prefix only when a turn is selected for deletion. Token-level deletion breaks it at every call with an eligible turn.

Depending on the provider, a cached token costs 0.05 to 0.5 times the price of a new input token~\cite{anthropicpricing,openaipricing,deepseekpricing,googlepricing}.
We pick a price ratio of $1 : 0.1 : 5$ for input, cached input, and output tokens.
The cached-input ratio of 0.1 lies within this range.
The caching-aware model ignores cache-write premiums and assumes that every matching prefix is served from the cache, so its cost estimate is a lower bound on realistic costs.

\subsection{Statistical Analysis}
\label{sec:design-stats}

\noindent
The reference for every comparison is the five uncompressed runs of the same agent.
For tasks solved and the re-fetch rate, we compare a configuration with the per-task mean of the five runs.
Run length and cost have long-tailed distributions, including capped runs, so we summarize these metrics with medians.
For these two metrics, we take the median of the configuration over the 100 tasks, divide it by the median of each uncompressed run, and average the five ratios.

We make two kinds of comparison.
The first is a configuration against the reference, as above.
The second compares two configurations that differ in one dimension, on the same agent: delayed against immediate timing (RQ1 and RQ2), or turn-level against token-level granularity at the same scope, amount, and timing (RQ1).
Every comparison is paired by task, and its uncertainty is a paired bootstrap confidence interval over the 100 tasks.
We resample the 100 tasks with replacement 10,000 times.
On each resample, we compute the statistic of the configuration on the drawn tasks and the statistic of the reference on the same drawn tasks (the mean task-solving outcome, or the median of turns or cost), and then their change or ratio.
The 2.5th and 97.5th percentiles of these 10,000 values form the 95\% interval.
A change is significant when its interval excludes zero (for a change in tasks solved) or one (for a ratio of turns, re-fetch rate, or cost).

\subsection{Implementation and Data}
\label{sec:design-impl}

\noindent
We conduct all experiments on a cluster with NVIDIA GH200 Hopper graphics processing units (GPUs), compute capability 9.0, and CUDA 12.8.
The experiments run in Apptainer containers with the vLLM~\cite{kwon2023:efficient} inference stack and incur no metered application programming interface (API) charges.
We serve the model with a context window of approximately 262,000 tokens, high reasoning effort, temperature zero, and a limit of 8,192 output tokens per call.
We keep full logs for all runs: prompts sent, generated tokens, and the original text of every deleted turn.
The replication package holds all 9,400 runs with their logs and the analysis code.

\section{RQ1: How Does Deletion Affect Task Solving?}
\label{sec:rq1}

\noindent
Table~\ref{tab:rq1} reports, for each configuration and agent, the number of tasks solved out of 100 and the change in median run length relative to the uncompressed baseline.
Bold marks a statistically significant change.
The table omits counts of capped runs, loops, and runs without a patch for readability.
The following analysis reports these counts where relevant.
The replication package includes all counts for every configuration.

\begin{table*}[t]\centering\footnotesize
\caption{RQ1: Tasks solved (of 100) and change in median run length. Bold marks statistically significant changes from the uncompressed baseline.
--: the median run reaches the 200-turn budget.}
\label{tab:rq1}
\setlength{\tabcolsep}{5pt}
\begin{tabular}{l l r @{\hspace{14pt}} r r @{\hspace{12pt}} r r @{\hspace{18pt}} r r @{\hspace{12pt}} r r}
  \toprule
   &  &  & \multicolumn{4}{c}{\textbf{Mini-SWE-agent}} & \multicolumn{4}{c}{\textbf{Trae agent}} \\
  \cmidrule(lr){4-7}\cmidrule(lr){8-11}
   &  &  & \multicolumn{2}{c}{Delayed} & \multicolumn{2}{c}{Immediate} & \multicolumn{2}{c}{Delayed} & \multicolumn{2}{c}{Immediate} \\
  \cmidrule(lr){4-5}\cmidrule(lr){6-7}\cmidrule(lr){8-9}\cmidrule(lr){10-11}
  Scope & Granularity & \% Deleted & Tasks solved & Turns & Tasks solved & Turns & Tasks solved & Turns & Tasks solved & Turns \\
  \midrule
  \multicolumn{3}{l}{Uncompressed} & 72.6 (71--76) & 31.5 & &  & 74.6 (74--76) & 42.5 & &  \\
  \midrule
  Reasoning & Turn & 25 & \resultcell{resultBetter!8}{71} & 0\% & \resultcell{resultBetter!8}{75} & +11\% & \resultcell{resultBetter!8}{70} & \textminus9\% & \resultcell{resultBetter!8}{76} & +1\% \\
   &  & 50 & \resultcell{resultBetter!8}{69} & +5\% & \resultcell{resultBetter!8}{73} & \resultcell{resultWorse!8.9}{\textbf{+13\%}} & \resultcell{resultBetter!8}{72} & \textminus3\% & \resultcell{resultWorse!10.4}{\textbf{69}} & \textminus5\% \\
   &  & 75 & \resultcell{resultBetter!8}{71} & +3\% & \resultcell{resultBetter!8}{71} & \resultcell{resultWorse!8.9}{\textbf{+13\%}} & \resultcell{resultBetter!8}{79} & \resultcell{resultBetter!8.7}{\textbf{\textminus11\%}} & \resultcell{resultBetter!8}{72} & \textminus6\% \\
   &  & 100 & \resultcell{resultBetter!8}{74} & +11\% & \resultcell{resultBetter!8}{75} & \resultcell{resultWorse!15.1}{\textbf{+105\%}} & \resultcell{resultBetter!8}{72} & \textminus4\% & \resultcell{resultBetter!8}{77} & \resultcell{resultWorse!10.4}{\textbf{+35\%}} \\
   & Token & 25 & \resultcell{resultBetter!8}{73} & 0\% & \resultcell{resultBetter!8}{74} & +6\% & \resultcell{resultBetter!8}{74} & \resultcell{resultBetter!8.7}{\textbf{\textminus10\%}} & \resultcell{resultBetter!8}{77} & \textminus7\% \\
   &  & 50 & \resultcell{resultBetter!8}{75} & 0\% & \resultcell{resultBetter!8}{74} & \resultcell{resultWorse!9.3}{\textbf{+19\%}} & \resultcell{resultBetter!8}{75} & \resultcell{resultBetter!9.0}{\textbf{\textminus15\%}} & \resultcell{resultBetter!8}{76} & +2\% \\
   &  & 75 & \resultcell{resultBetter!8}{74} & \textminus11\% & \resultcell{resultBetter!8}{71} & \resultcell{resultWorse!11.3}{\textbf{+49\%}} & \resultcell{resultBetter!8}{79} & \resultcell{resultBetter!8.7}{\textbf{\textminus11\%}} & \resultcell{resultBetter!8}{74} & \resultcell{resultWorse!9.0}{\textbf{+15\%}} \\
  \midrule
  Tool output & Turn & 25 & \resultcell{resultBetter!8}{72} & +3\% & \resultcell{resultBetter!8}{75} & \resultcell{resultWorse!9.7}{\textbf{+25\%}} & \resultcell{resultBetter!8}{74} & +3\% & \resultcell{resultBetter!8}{73} & \resultcell{resultWorse!9.9}{\textbf{+29\%}} \\
   &  & 50 & \resultcell{resultBetter!8}{73} & \resultcell{resultWorse!9.4}{\textbf{+21\%}} & \resultcell{resultBetter!8}{72} & \resultcell{resultWorse!13.8}{\textbf{+86\%}} & \resultcell{resultBetter!8}{72} & \resultcell{resultWorse!9.3}{\textbf{+19\%}} & \resultcell{resultBetter!8}{73} & \resultcell{resultWorse!15.5}{\textbf{+111\%}} \\
   &  & 75 & \resultcell{resultBetter!8}{70} & \resultcell{resultWorse!8.9}{\textbf{+14\%}} & \resultcell{resultWorse!15.1}{\textbf{56}} & \resultcell{resultWorse!22.2}{\textbf{+211\%}} & \resultcell{resultBetter!8}{70} & \resultcell{resultWorse!10.4}{\textbf{+36\%}} & \resultcell{resultWorse!11.7}{\textbf{66}} & \resultcell{resultWorse!31.5}{\textbf{+350\%}} \\
   &  & 100 & \resultcell{resultBetter!8}{69} & \resultcell{resultWorse!9.3}{\textbf{+19\%}} & \resultcell{resultWorse!33.1}{\textbf{14}} & \resultcell{resultWorse!40.0}{\textbf{+476\%}} & \resultcell{resultBetter!8}{70} & \resultcell{resultWorse!12.9}{\textbf{+73\%}} & \resultcell{resultWorse!40.0}{\textbf{0}} & -- \\
   & Token & 25 & \resultcell{resultBetter!8}{71} & +3\% & \resultcell{resultBetter!8}{71} & \resultcell{resultWorse!12.4}{\textbf{+65\%}} & \resultcell{resultBetter!8}{74} & \resultcell{resultWorse!9.1}{\textbf{+16\%}} & \resultcell{resultWorse!18.6}{\textbf{50}} & \resultcell{resultWorse!21.7}{\textbf{+204\%}} \\
   &  & 50 & \resultcell{resultBetter!8}{69} & \resultcell{resultWorse!9.5}{\textbf{+22\%}} & \resultcell{resultWorse!11.3}{\textbf{65}} & \resultcell{resultWorse!19.7}{\textbf{+174\%}} & \resultcell{resultBetter!8}{70} & \resultcell{resultWorse!10.6}{\textbf{+39\%}} & \resultcell{resultWorse!27.6}{\textbf{29}} & -- \\
   &  & 75 & \resultcell{resultBetter!8}{70} & \resultcell{resultWorse!10.4}{\textbf{+35\%}} & \resultcell{resultWorse!27.6}{\textbf{27}} & -- & \resultcell{resultWorse!11.3}{\textbf{67}} & \resultcell{resultWorse!11.0}{\textbf{+45\%}} & \resultcell{resultWorse!37.9}{\textbf{5}} & -- \\
  \midrule
  Both & Turn & 25 & \resultcell{resultBetter!8}{72} & +3\% & \resultcell{resultBetter!8}{72} & \resultcell{resultWorse!9.6}{\textbf{+24\%}} & \resultcell{resultBetter!8}{72} & +15\% & \resultcell{resultBetter!8}{69} & \resultcell{resultWorse!10.4}{\textbf{+36\%}} \\
   &  & 50 & \resultcell{resultBetter!8}{74} & +3\% & \resultcell{resultWorse!10.8}{\textbf{66}} & \resultcell{resultWorse!12.6}{\textbf{+68\%}} & \resultcell{resultBetter!8}{72} & \resultcell{resultWorse!9.8}{\textbf{+27\%}} & \resultcell{resultBetter!8}{74} & \resultcell{resultWorse!14.9}{\textbf{+102\%}} \\
   &  & 75 & \resultcell{resultBetter!8}{70} & \resultcell{resultWorse!9.7}{\textbf{+25\%}} & \resultcell{resultBetter!8}{67} & \resultcell{resultWorse!24.4}{\textbf{+244\%}} & \resultcell{resultBetter!8}{69} & \resultcell{resultWorse!14.2}{\textbf{+92\%}} & \resultcell{resultBetter!8}{72} & \resultcell{resultWorse!30.7}{\textbf{+338\%}} \\
   &  & 100 & \resultcell{resultWorse!13.8}{\textbf{59}} & \resultcell{resultWorse!14.6}{\textbf{+98\%}} & \resultcell{resultWorse!38.3}{\textbf{2}} & \resultcell{resultWorse!37.1}{\textbf{+433\%}} & \resultcell{resultWorse!21.6}{\textbf{43}} & -- & \resultcell{resultWorse!40.0}{\textbf{0}} & -- \\
   & Token & 25 & \resultcell{resultBetter!8}{69} & +16\% & \resultcell{resultWorse!25.8}{\textbf{31}} & +21\% & \resultcell{resultBetter!8}{71} & \resultcell{resultWorse!8.8}{\textbf{+12\%}} & \resultcell{resultWorse!15.1}{\textbf{58}} & -- \\
   &  & 50 & \resultcell{resultWorse!10.8}{\textbf{66}} & \resultcell{resultWorse!9.7}{\textbf{+25\%}} & \resultcell{resultWorse!31.8}{\textbf{17}} & \resultcell{resultWorse!11.5}{\textbf{+52\%}} & \resultcell{resultBetter!8}{77} & \resultcell{resultWorse!15.1}{\textbf{+106\%}} & \resultcell{resultWorse!25.4}{\textbf{34}} & -- \\
   &  & 75 & \resultcell{resultBetter!8}{68} & \resultcell{resultWorse!10.2}{\textbf{+32\%}} & \resultcell{resultWorse!33.1}{\textbf{14}} & \resultcell{resultWorse!33.6}{\textbf{+381\%}} & \resultcell{resultWorse!14.3}{\textbf{60}} & -- & \resultcell{resultWorse!37.9}{\textbf{5}} & -- \\
  \bottomrule
\end{tabular}
  \vspace*{-1ex}
\end{table*}

\subsection{Delayed Deletion: Scope, Granularity, and Amount}

\noindent
\textbf{Tasks solved.}
Only five of the 42 delayed-configuration evaluations (in bold) show a significant loss, while the confidence intervals for the other 37 include zero change.
Three of these five solve more than ten fewer tasks than the baseline,
and all three delete reasoning and tool output together: Mini-SWE-agent loses 14 tasks and Trae agent 32 when every turn is deleted whole, and Trae agent loses 15 under token-level deletion of 75\% of both.
The other two drops are seven tasks for Mini-SWE-agent with 50\% token-level deletion of both scopes and eight for Trae agent with 75\% token-level deletion of tool output.

\textbf{Scope.}
Delayed reasoning deletion shows no statistically significant decrease in tasks solved or increase in run length.
Across the 14 evaluations of reasoning deletion, the change in tasks solved ranges from $-5$ to $+4$ relative to the baseline. Some Trae agent configurations reduce median run length by up to 15\%.
At a fixed deletion percentage, reasoning deletion removes fewer tokens than deletion of tool output or both scopes (Section~\ref{sec:threats}). The experiment therefore does not isolate the effect of scope from the number of tokens removed.

\textbf{Granularity.}
Granularity primarily affects run length in these comparisons.
Only one pair differs significantly in tasks solved by the paired bootstrap: Trae agent, both, 75\%, where turn-level deletion solves 69 tasks and token-level deletion 60.
Granularity also affects run length.
Token-level deletion produces longer runs than turn-level deletion in ten of the twelve pairs that delete tool output. The difference increases with the deletion percentage.
At 75\% deletion of both scopes, Trae agent's median run reaches the turn cap under token-level deletion, compared with a 92\% increase over the baseline under turn-level deletion.
In the six reasoning pairs, runs are equal or shorter under token-level deletion.

\textbf{Amount.}
Up to 50\%, only one evaluation (Mini-SWE-agent) loses tasks significantly.
At 75\%, three evaluations on Trae agent solve six to 15 fewer tasks, all involving tool output deletion.
Deleting 100\% of both reasoning and tool output significantly reduces tasks solved for both agents.
Increased deletion is also associated with longer runs.
Of the 28 evaluations that delete tool output, 21 have significantly longer runs, compared with none of the 14 evaluations that delete reasoning alone.
Under turn-level deletion of both scopes, Trae agent’s median run grows from 15\% above the baseline at 25\% deletion to the cap at 100\%.
The number of capped runs increases from four to 92 out of 100, whereas tasks solved decrease significantly only at 100\% deletion, from a baseline mean of 74.6 to 43.
Longer runs also occur without a statistically significant decrease in tasks solved.
Under turn-level 100\% deletion of tool output, Trae agent solves 70 tasks, compared with a baseline mean of 74.6, while median run length increases by 73\% and 29 runs reach the cap.
Reporting only tasks solved thus hides longer runs and budget exhaustion.

\textbf{Comparison with AgentDiet.}
AgentDiet~\cite{xiao2026:reducing} evaluates two fixed baselines on Trae agent, on the same 100 tasks and with the same two-turn delay: one deletes 75\% of the tokens of a turn at random and the other deletes the turn entirely, both applied to reasoning and tool output together~\cite{xiao2026:reducing}.
They correspond to our token-level deletion of 75\% of both and our turn-level deletion of 100\% of both.
Both studies agree that deleting the whole turn costs more tasks and more turns than deleting 75\% of its tokens.
The magnitudes differ: AgentDiet reports 1 and 7 fewer tasks solved against our 15 and 32, and 13\% to 14\% more turns where our runs reach the cap.
Three design differences limit direct comparison: AgentDiet rewrites only turns longer than 500 tokens rather than every turn, uses Claude 4 Sonnet rather than DeepSeek-V4-Flash, and limits runs to 50 turns rather than 200.
\begin{myframed}
\noindent
\textbf{Answer to RQ1: Delayed deletion.}
With delayed deletion, 37 of 42 evaluations show no statistically significant decrease in tasks solved.
However, runs lengthen significantly in 21, all deleting tool output.
\end{myframed}

\noindent\textbf{Implications for future work.}
Our results call for evaluations of trajectory reduction that measure tasks solved, execution length, and budget exhaustion under common turn limits, since agents can preserve effectiveness through additional work.
The contrast between reasoning and tool output deletion also motivates memory-management policies that treat these content types differently.
Future comparisons should match the number of tokens removed to determine whether such policies improve the tradeoff between effectiveness and execution effort.

\subsection{Immediate Deletion: The Effect of Timing}

\noindent
Without the delay, a deleted output is rewritten before the first call that would show it, so the agent never sees it intact.
At 100\% of tool output or of both, the agent therefore never sees any tool output at all.
Immediate deletion raises significant losses in tasks solved (in bold) from five to 19 of the 42 evaluations per timing, and significantly longer runs from 21 to 34.
No immediate configuration has significantly shorter runs than the baseline.
Paired bootstrap comparisons of configurations that differ only in timing show significantly fewer tasks solved with immediate deletion in 16 of the 42 comparisons.
The two agents fail differently.
In the eight Trae agent configurations with fewer than 60 tasks solved, 33 to 100 of the runs loop, against two to five in the uncompressed runs.
In the six Mini-SWE-agent configurations with fewer than 35 tasks solved, 38 to 94 of the runs end without a patch, and Mini-SWE-agent loops in at most one run in any configuration.

\textbf{Turn-level deletion.}
At 25\% and 50\%, no turn-level configuration differs significantly from its delayed value, although runs are up to twice as long as the baseline when deletion affects tool output.
Two of these configurations, Mini-SWE-agent with both and Trae agent with reasoning deleted at 50\%, are about six tasks below the reference.
At 75\% deletion, the effects differ between agents.
Mini-SWE-agent loses 14 tasks when tool output is deleted, while Trae agent keeps its delayed value but loops in 41 to 60 of its runs, with a median run close to the cap.
At 100\% deletion of tool output or both scopes,
Trae agent solves no tasks, and with both deleted, it repeats one directory listing until the cap in every run.
Mini-SWE-agent solves two tasks with both deleted and 14 with tool output deleted, without ever observing file contents.
Its 14 successful patches warrant further analysis to distinguish reliance on issue descriptions, memorized fixes, and other explanations.

\textbf{Token-level deletion.}
Immediate token-level deletion of tool output reduces Trae agent’s tasks solved at every deletion percentage, to five at 75\%, with most runs capped from 50\% onward.
Mini-SWE-agent shows no significant difference from delayed deletion up to 50\% deletion and solves 27 tasks at 75\% deletion.
With both scopes deleted, Mini-SWE-agent solves 31 tasks at 25\% against 69 under delayed deletion.
Of the immediate runs, 45 end without a patch, and none reach the cap.
Compared with delayed deletion, Trae agent solves 13 fewer tasks at 25\% deletion.
It solves 34 and five tasks at 50\% and 75\% deletion, respectively.
The content left by each granularity offers an interpretation of this difference, drawn from Trae agent's traces and not an established cause.
Turn-level deletion replaces the removed content with an explicit deletion notice, whereas token-level deletion leaves fragments that the agent sometimes reads as repository corruption and attempts to repair.
Loop counts are consistent with this interpretation.
At 50\% of tool output, 67 of Trae agent's runs loop under token-level deletion against eight under turn-level deletion, compared with nine and six looping runs under delayed deletion and two to five in the uncompressed baseline.

\textbf{Reasoning deletion.}
Immediate and delayed reasoning deletion yield similar numbers of tasks solved. Confidence intervals include zero change in 13 of the 14 evaluations.
The exception
solves six fewer tasks than the baseline.
Run length shows a different pattern.
No delayed reasoning evaluation has significantly longer runs, but seven of the 14 immediate ones do, by under 20\% in all but three.
Mini-SWE-agent's runs are 49\% longer under token-level deletion of 75\%, and turn-level deletion of 100\% lengthens Trae agent's runs by 35\% and Mini-SWE-agent's by 105\%.
At 75\% deletion, the corresponding increase is 13\% for Mini-SWE-agent, with no significant change for Trae agent.

\begin{myframed}
\noindent
\textbf{Answer to RQ1: Immediate deletion.}
Without the delay, 19 of 42 evaluations show significantly fewer tasks solved than the baseline, and 34 have significantly longer runs.
While reasoning deletion and turn-level deletion of up to 50\% of tool output generally retain their delayed effectiveness, token-level deletion of tool output does not.
\end{myframed}

\noindent\textbf{Implications for future work.}
The timing comparisons motivate studies of memory management that jointly vary retention delay and deletion scope.
Such studies should test whether reasoning and tool outputs require different protection periods and measure tasks solved, execution length, and failure behavior under each policy.
Analyses of loops, missing patches, and budget exhaustion would clarify how retention choices affect execution beyond the final task outcome.

\section{RQ2: How Does the Agent Compensate for Deleted Content?}
\label{sec:rq2}

\noindent
Table~\ref{tab:rq2} gives, for each configuration and agent, the re-fetch rate as a ratio to the uncompressed runs, where 19 of every 100 reads are re-fetches for Mini-SWE-agent and 9 for Trae agent.
A ratio of 2.0 indicates twice the baseline re-fetch rate.
Bold marks a statistically significant difference from the baseline.

\begin{table}[t]\centering\scriptsize
\caption{RQ2 and RQ3: Re-fetch rates under delayed and immediate deletion, and costs under delayed deletion with flat and caching-aware pricing, as ratios to the uncompressed runs.}
\label{tab:rq2}\label{tab:rq3}
\setlength{\tabcolsep}{1.5pt}
\begin{tabular}{l l r r r p{0pt} r r p{1pt} r r p{0pt} r r}
  \toprule
  & & & \multicolumn{5}{c}{\textbf{Mini-SWE-agent}} & & \multicolumn{5}{c}{\textbf{Trae agent}} \\
  \cmidrule(lr){4-8}\cmidrule(lr){10-14}
  & & & \multicolumn{2}{c}{Re-fetch} & & \multicolumn{2}{c}{Cost} & & \multicolumn{2}{c}{Re-fetch} & & \multicolumn{2}{c}{Cost} \\
  \cmidrule(lr){4-5}\cmidrule(lr){7-8}\cmidrule(lr){10-11}\cmidrule(lr){13-14}
  Scope & Granul. & Del\% & Delay & Imm. & & Flat & Cache & & Delay & Imm. & & Flat & Cache \\
  \midrule
  Reasoning & Turn & 25 & 1.0 & 1.0 &  & 0.92 & 1.12 &  & 0.9 & 1.0 &  & 0.90 & 1.05 \\
   &  & 50 & 1.0 & 1.1 &  & 0.93 & 1.06 &  & 1.0 & 0.9 &  & 0.85 & 1.11 \\
   &  & 75 & \resultcell{resultWorse!9.6}{\textbf{1.2}} & \resultcell{resultWorse!8.8}{\textbf{1.1}} &  & 0.84 & 1.23 &  & 1.0 & \resultcell{resultWorse!9.6}{\textbf{1.2}} &  & \resultcell{resultBetter!9.6}{\textbf{0.80}} & 1.11 \\
   &  & 100 & \resultcell{resultWorse!8.8}{\textbf{1.1}} & \resultcell{resultWorse!15.3}{\textbf{1.9}} &  & 0.99 & \resultcell{resultWorse!11.6}{\textbf{1.44}} &  & 0.9 & \resultcell{resultWorse!15.3}{\textbf{1.9}} &  & \resultcell{resultBetter!9.5}{\textbf{0.82}} & \resultcell{resultWorse!9.8}{\textbf{1.22}} \\
   & Token & 25 & 1.0 & 1.0 &  & 1.05 & \resultcell{resultWorse!10.8}{\textbf{1.35}} &  & \resultcell{resultBetter!10.4}{\textbf{0.7}} & 0.9 &  & 0.84 & \resultcell{resultWorse!9.3}{\textbf{1.16}} \\
   &  & 50 & 1.1 & 1.1 &  & 0.87 & 1.24 &  & \resultcell{resultBetter!9.6}{\textbf{0.8}} & 1.2 &  & \resultcell{resultBetter!10.1}{\textbf{0.74}} & 1.09 \\
   &  & 75 & 1.1 & \resultcell{resultWorse!12.1}{\textbf{1.5}} &  & 0.84 & \resultcell{resultWorse!11.9}{\textbf{1.48}} &  & 1.0 & \resultcell{resultWorse!12.1}{\textbf{1.5}} &  & \resultcell{resultBetter!10.1}{\textbf{0.74}} & \resultcell{resultWorse!9.5}{\textbf{1.19}} \\
  \midrule
  Tool output & Turn & 25 & \resultcell{resultWorse!9.6}{\textbf{1.2}} & \resultcell{resultWorse!9.6}{\textbf{1.2}} &  & 0.97 & 1.03 &  & \resultcell{resultWorse!12.1}{\textbf{1.5}} & \resultcell{resultWorse!16.9}{\textbf{2.1}} &  & 0.88 & 1.04 \\
   &  & 50 & \resultcell{resultWorse!10.4}{\textbf{1.3}} & \resultcell{resultWorse!12.1}{\textbf{1.5}} &  & 0.98 & 1.11 &  & \resultcell{resultWorse!16.9}{\textbf{2.1}} & \resultcell{resultWorse!25.1}{\textbf{3.1}} &  & 0.80 & 1.09 \\
   &  & 75 & \resultcell{resultWorse!12.1}{\textbf{1.5}} & \resultcell{resultWorse!12.1}{\textbf{1.5}} &  & 0.85 & 1.19 &  & \resultcell{resultWorse!23.4}{\textbf{2.9}} & \resultcell{resultWorse!40.5}{\textbf{5.0}} &  & 0.92 & \resultcell{resultWorse!11.0}{\textbf{1.37}} \\
   &  & 100 & \resultcell{resultWorse!12.9}{\textbf{1.6}} & 1.0 &  & \resultcell{resultBetter!10.4}{\textbf{0.71}} & \resultcell{resultWorse!9.6}{\textbf{1.20}} &  & \resultcell{resultWorse!37.2}{\textbf{4.6}} & \resultcell{resultWorse!43.8}{\textbf{5.4}} &  & 1.01 & \resultcell{resultWorse!13.4}{\textbf{1.67}} \\
   & Token & 25 & \resultcell{resultWorse!11.2}{\textbf{1.4}} & \resultcell{resultWorse!11.2}{\textbf{1.4}} &  & 0.94 & \resultcell{resultWorse!9.8}{\textbf{1.22}} &  & \resultcell{resultWorse!12.9}{\textbf{1.6}} & \resultcell{resultWorse!30.8}{\textbf{3.8}} &  & 1.10 & \resultcell{resultWorse!11.9}{\textbf{1.48}} \\
   &  & 50 & \resultcell{resultWorse!12.9}{\textbf{1.6}} & \resultcell{resultWorse!12.1}{\textbf{1.5}} &  & 1.07 & \resultcell{resultWorse!11.5}{\textbf{1.43}} &  & \resultcell{resultWorse!21.8}{\textbf{2.7}} & \resultcell{resultWorse!41.3}{\textbf{5.1}} &  & \resultcell{resultWorse!10.1}{\textbf{1.26}} & \resultcell{resultWorse!13.6}{\textbf{1.69}} \\
   &  & 75 & \resultcell{resultWorse!13.7}{\textbf{1.7}} & \resultcell{resultWorse!12.9}{\textbf{1.6}} &  & 1.04 & \resultcell{resultWorse!11.9}{\textbf{1.48}} &  & \resultcell{resultWorse!29.9}{\textbf{3.7}} & \resultcell{resultWorse!56.8}{\textbf{7.0}} &  & 1.05 & \resultcell{resultWorse!13.7}{\textbf{1.70}} \\
  \midrule
  Both & Turn & 25 & \resultcell{resultWorse!9.6}{\textbf{1.2}} & \resultcell{resultWorse!10.4}{\textbf{1.3}} &  & 1.05 & 1.14 &  & \resultcell{resultWorse!13.7}{\textbf{1.7}} & \resultcell{resultWorse!21.8}{\textbf{2.7}} &  & 1.03 & 1.27 \\
   &  & 50 & \resultcell{resultWorse!10.4}{\textbf{1.3}} & \resultcell{resultWorse!12.9}{\textbf{1.6}} &  & \resultcell{resultBetter!9.7}{\textbf{0.79}} & 1.03 &  & \resultcell{resultWorse!20.2}{\textbf{2.5}} & \resultcell{resultWorse!35.6}{\textbf{4.4}} &  & 0.97 & \resultcell{resultWorse!10.5}{\textbf{1.31}} \\
   &  & 75 & \resultcell{resultWorse!13.7}{\textbf{1.7}} & \resultcell{resultWorse!16.9}{\textbf{2.1}} &  & 0.86 & \resultcell{resultWorse!10.5}{\textbf{1.31}} &  & \resultcell{resultWorse!32.4}{\textbf{4.0}} & \resultcell{resultWorse!52.7}{\textbf{6.5}} &  & 1.13 & \resultcell{resultWorse!14.0}{\textbf{1.74}} \\
   &  & 100 & \resultcell{resultWorse!19.4}{\textbf{2.4}} & \resultcell{resultWorse!17.8}{\textbf{2.2}} &  & 0.98 & \resultcell{resultWorse!15.9}{\textbf{1.97}} &  & \resultcell{resultWorse!60.8}{\textbf{7.5}} & \resultcell{resultWorse!86.0}{\textbf{10.6}} &  & 1.12 & \resultcell{resultWorse!27.2}{\textbf{3.36}} \\
   & Token & 25 & \resultcell{resultWorse!9.6}{\textbf{1.2}} & \resultcell{resultWorse!10.4}{\textbf{1.3}} &  & 1.11 & \resultcell{resultWorse!12.5}{\textbf{1.56}} &  & \resultcell{resultWorse!19.4}{\textbf{2.4}} & \resultcell{resultWorse!36.4}{\textbf{4.5}} &  & 0.97 & \resultcell{resultWorse!11.5}{\textbf{1.43}} \\
   &  & 50 & \resultcell{resultWorse!12.1}{\textbf{1.5}} & \resultcell{resultWorse!14.5}{\textbf{1.8}} &  & 1.08 & \resultcell{resultWorse!11.7}{\textbf{1.46}} &  & \resultcell{resultWorse!31.6}{\textbf{3.9}} & \resultcell{resultWorse!71.4}{\textbf{8.8}} &  & \resultcell{resultWorse!13.0}{\textbf{1.61}} & \resultcell{resultWorse!18.6}{\textbf{2.31}} \\
   &  & 75 & \resultcell{resultWorse!16.1}{\textbf{2.0}} & \resultcell{resultWorse!19.4}{\textbf{2.4}} &  & 1.10 & \resultcell{resultWorse!16.2}{\textbf{2.01}} &  & \resultcell{resultWorse!51.1}{\textbf{6.3}} & \resultcell{resultWorse!75.4}{\textbf{9.3}} &  & \resultcell{resultWorse!30.2}{\textbf{3.73}} & \resultcell{resultWorse!40.0}{\textbf{4.94}} \\
  \bottomrule
\end{tabular}
\end{table}

\textbf{Scope.}
Deletion scope affects re-fetching similarly to the pattern for run length in Section~\ref{sec:rq1}.
Deleting both scopes triggers the most re-fetching, followed by tool output, then reasoning. At matched amounts and granularities, deleting both scopes yields higher re-fetch ratios than deleting tool output in all Trae agent comparisons and most Mini-SWE-agent comparisons.

\textbf{Granularity.}
Agents re-fetch more under token-level deletion than under turn-level deletion in 21 of the 24 pairs that delete tool output or both, at the same scope and amount, and the other three are ties.

\textbf{Amount.}
Section~\ref{sec:rq1} shows that run length increases with the percentage of tool output deleted.
The additional turns include repository exploration, testing, and editing.
Table~\ref{tab:rq2} identifies re-fetching as one source of additional work that increases when deleting tool output.
Under delayed turn-level deletion, Trae agent’s ratio grows from 1.7× at 25\% to 7.5× at 100\% for both, and from 1.5× to 4.6× for tool output.
The corresponding ratios at 100\% deletion for Mini-SWE-agent are $2.4\times$ and $1.6\times$, respectively.
Of the 28 evaluations that delete tool output at each timing, all 28 show a significant increase with the delay and 27 without it.
The exception, Mini-SWE-agent at 100\% of tool output without the delay (1.0×), never receives any tool output and explores unread content rather than re-reading: in its 41 capped runs, 36\% of all calls read content not seen before, 35\% are scripts, and 7\% re-read content already seen.

\textbf{Timing.}
Immediate deletion produces significantly higher re-fetch rates than delayed deletion in 25 of the 42 paired bootstrap comparisons, 18 of 21 for Trae agent and seven of 21 for Mini-SWE-agent.
The increase is largest in the token-level rows.
Trae agent goes from 3.7× with the delay to 7.0× without it at 75\% of tool output, and from 3.9× to 8.8× at 50\% of both.
For Mini-SWE-agent, tool output re-fetch ratios differ little between timings below 100\% deletion.
The highest value in the table, $10.6\times$, is Trae agent under immediate turn-level deletion of both at 100\%.

\textbf{Reasoning.}
Delayed reasoning deletion produces no consistent increase in re-fetching.
Re-fetch ratios range from 0.7 to 1.2 across the 14 evaluations, and the four significant differences from the baseline are small, point in opposite directions
(two below one, two above one),
and show no trend with deletion percentage.
Two immediate configurations stand out.
In both agents, turn-level deletion of 100\% reaches $1.9\times$, and token-level deletion of 75\% reaches $1.5\times$.
Under both granularities, removing the latest reasoning increases re-fetching; removing reasoning older than two turns does not.

\textbf{Deleted versus retained outputs.}
A higher re-fetch rate can mean that the agent retrieves deleted content again or that it retrieves both deleted and retained content more.
To distinguish these explanations, we compare subsequent retrieval of deleted and retained outputs within each run.
Under turn-level deletion, independent random draws determine retention.
Assignment therefore does not systematically select outputs by age, file, or position, supporting interpretation of differences in subsequent retrieval as responses to deletion.
Table~\ref{tab:rq2fb} gives the percentage of kept and of deleted reads whose output is re-fetched later, for the turn-level configurations that delete tool output or both at 25\% to 75\%.

\begin{table}[t]\centering\footnotesize
\caption{RQ2: Percentages of retained and deleted file-read outputs subsequently re-fetched under turn-level deletion. Each cell lists the retained-output percentage followed by the deleted-output percentage.}
\label{tab:rq2fb}
\setlength{\tabcolsep}{3pt}
\begin{tabular}{l r @{\hspace{4pt}} r r @{\hspace{6pt}} r r}
  \toprule
  & & \multicolumn{2}{c}{\textbf{Mini-SWE-agent}} & \multicolumn{2}{c}{\textbf{Trae agent}} \\
  \cmidrule(lr){3-4}\cmidrule(lr){5-6}
  Scope & \% Deleted & Delayed & Immediate & Delayed & Immediate \\
  \midrule
  \multicolumn{2}{l}{Uncompressed} & 13\% & & 9\% & \\
  \midrule
  Tool output & 25 & 11\% / 22\% & 14\% / 30\% & 9\% / 28\% & 9\% / 45\% \\
   & 50 & 12\% / 18\% & 10\% / 25\% & 9\% / 27\% & 7\% / 36\% \\
   & 75 & 11\% / 17\% & 8\% / 16\% & 7\% / 21\% & 7\% / 28\% \\
  \midrule
  Both & 25 & 12\% / 23\% & 12\% / 28\% & 9\% / 37\% & 10\% / 50\% \\
   & 50 & 11\% / 22\% & 12\% / 27\% & 9\% / 27\% & 7\% / 38\% \\
   & 75 & 8\% / 19\% & 9\% / 19\% & 8\% / 24\% & 6\% / 24\% \\
  \bottomrule
\end{tabular}
\end{table}

In every evaluated configuration, kept outputs are re-fetched about as often as in the uncompressed runs.
The extra re-fetches are of the deleted content.
Deleted outputs are re-fetched 1.5 to five times as often as kept ones, and
the paired bootstrap analysis finds the difference significant in all but one of the 24 evaluations, Mini-SWE-agent at 50\% of tool output with the delay.
The deletion configurations do not distinguish a needed output from an unneeded one, so some of what they delete is content the agent still needs.
The comparison supports selective retrieval of deleted content rather than a general increase in retrieval of retained content.
This pattern is consistent with the longer runs and mostly preserved effectiveness under turn-level tool output deletion in Section~\ref{sec:rq1}.

\textbf{The shape of a run.}
Tool calls per turn range from 0.92 to 1.12 times the baseline across all 84 evaluations.
Additional re-fetches therefore primarily appear as additional turns.
Deletion scope also affects reasoning length, measured in words per turn.
Deleting tool output shortens the reasoning.
Immediate tool output deletion at the two highest amounts shortens it by about 50\% while run lengths reach four times the baseline.
Under delayed tool output deletion, reasoning length remains close to the baseline.
Reasoning deletion lengthens it to twice the baseline with the delay, and five times without it, with turn counts below 100\% deletion staying close to the baseline.
The effect of deleting both scopes depends on the granularity.
Immediate turn-level deletion of 100\% shortens reasoning by over 50\%, similar to tool output deletion.
Token-level deletion raises reasoning per turn to as much as seven times the baseline for Mini-SWE-agent and four times for Trae agent, alongside run lengths of three to four times the baseline.
The reasoning length of every configuration is in the replication package.

\textbf{How far back the agent reaches.}
Lastly, we measure on the uncompressed runs how far back an agent reaches for content.
We exclude the 19 Trae agent runs that loop (Section~\ref{sec:design-metrics}), since their repeated reads place every re-fetch one turn apart.
Re-fetches tend to reach further back than edits.
The median interval between a read and its re-fetch is twelve turns for Trae agent and six for Mini-SWE-agent, whereas an edit follows the last read of the file by a median of four turns and one turn.
A two-turn delay therefore leaves 40\% of Trae agent's edits and 68\% of Mini-SWE-agent's with their last read intact, but only 9\% and 23\% of their re-fetches.
A ten-turn delay would leave 69\% and 92\% of the edits and 44\% and 66\% of the re-fetches intact.
These percentages describe the uncompressed agents, and since deletion may change behavior, they do not predict the effect of a longer delay under deletion.
They show that retrieval reaches well beyond the two-turn window, motivating evaluation of longer retention windows.

\begin{myframed}
\noindent
\textbf{Answer to RQ2.}
Tool output deletion raises re-fetch rates up to $7.5\times$ with the delay and $10.6\times$ without it. The extra re-fetches are of the deleted content, while re-fetching of retained content stays at the uncompressed baseline.
Delayed reasoning deletion produces no consistent increase, and immediate reasoning deletion increases re-fetching only at the highest percentages.
The extra re-fetches appear as turns, and they reach back a median six to twelve turns, well beyond the two-turn delay.
\end{myframed}

\noindent\textbf{Implications for future work.}
Future deletion studies should distinguish dependence on retained context from the ability to recover missing information.
Selective re-fetching shows why successful execution after deletion does not establish that the removed content is unnecessary.
This finding calls for joint evaluation of retention and retrieval, including what remains accessible and the effort required to recover it.
Trace analyses should connect repeated reads to the content deleted and subsequent actions, distinguishing recovery of missing observations from unproductive loops.
Since re-fetches reach well beyond the two-turn delay, longer retention windows and memory-management approaches that coordinate what stays in the prompt with what remains retrievable are worth evaluating.

\section{RQ3: How Does Deletion Change Cost?}

\label{sec:rq3}
\noindent
Table~\ref{tab:rq3} reports, for each delayed configuration and agent, estimated cost relative to the uncompressed baseline under flat and caching-aware pricing (statistically significant differences in bold).
Ratios of 0.80 and 1.20 indicate costs 20\% below and above the baseline.
The table focuses on delayed configurations because most evaluations preserve tasks solved without a statistically significant decrease.
Immediate configurations produce larger decreases,
which complicates interpretation of per-run cost.
The timing paragraph below reports their cost results, while analysis of cost per task solved remains future work.
All data is available in the replication package.

\subsection{Cost Under Flat and Caching-Aware Pricing}

\noindent
\textbf{Pricing.}
Under flat pricing, estimated costs are below the baseline in 26 of 42 evaluations,
six showing significant cost reductions
and three being significantly more expensive.
Under caching-aware pricing, every evaluation has an estimated cost ratio above one,
significantly in 26 of 42.
The configurations closest to parity have estimated costs no more than 5\% above the uncompressed baseline.
Three of the six significant flat-pricing savings become significant increases, among them Mini-SWE-agent under turn-level deletion of all tool output, approximately 30\% below the baseline under flat pricing and 20\% above it under caching-aware pricing.
None of the six shows a significant decrease in tasks solved.

\textbf{Scope.}
Reasoning deletion accounts for four of the significant flat-pricing savings, all four of Trae agent’s, with estimated costs up to 26\% below the baseline.
Under caching-aware pricing, all 14 of its ratios exceed one, six significantly.
Tool output deletion produces one significantly lower cost under flat pricing, whereas nine of its 14 evaluations have significantly higher costs under caching-aware pricing.
Deleting both scopes produces the largest increase.
Under caching-aware pricing, 11 of its 14 evaluations are significantly more expensive, reaching approximately five times the uncompressed cost for Trae agent under token-level deletion of 75\%, whose median run hits the cap (Section~\ref{sec:rq1}).

\textbf{Granularity.}
Cost differences between granularities follow the pattern observed for run length.
Under caching-aware pricing, turn-level deletion costs no more than token-level deletion in 17 of the 18 matched pairs. The remaining pair differs by 0.02 in cost ratio.

\textbf{Timing.}
Without the delay, one evaluation is cheaper under flat pricing and none under caching-aware pricing:
Trae agent under turn-level deletion of 100\% of both scopes, for which every run repeatedly lists the root directory until reaching the cap, sending approximately half as many tokens as the baseline and solving no tasks.
Most evaluations of immediate configurations are significantly more expensive, 29 of the 42 evaluations under flat pricing and 32 under caching-aware pricing.
Immediate reasoning deletion mostly preserves tasks solved, supporting cost comparisons with delayed reasoning deletion.
None of the 14 evaluations is cheaper under either pricing model.
Under caching-aware pricing, all 14 ratios exceed one and seven increases are significant.
Immediate turn-level deletion of all reasoning reaches 1.7 times the baseline cost for Trae agent and 4 times the baseline cost for Mini-SWE-agent.

\subsection{Source of the Cost Difference}

\noindent
\textbf{Token volume and billing.}
Token counts fall below the baseline in 31 of the 42 delayed evaluations, by up to one third.
Most of the other eleven use token-level deletion, where longer runs offset the tokens removed from each prompt.
Tokens billed at the full input price increase in all 42 evaluations, ranging from approximately 1.25 to nearly 10 times the baseline count.
Re-billed tokens account for this increase (Section~\ref{sec:design-metrics}).
Every rewrite re-bills the prompt from the first changed token onwards.
Flat pricing charges all prompt tokens at the input price and does not represent changes in cache reuse.
Caching-aware pricing accounts for both the loss of cache discounts and additional tokens from longer runs.

\textbf{Cost components.}
The cost decomposition separates four components: cached input, new input, re-billed input, and output.
Re-billed tokens incur the new-input price but form a separate component to identify the effect of prefix invalidation.
Averaged over the five uncompressed runs, cached, new-input, and output tokens account for 50\%, 14\%, and 36\% of Mini-SWE-agent’s cost and 52\%, 14\%, and 34\% of Trae agent’s, with no re-billed cost in either baseline.

Deletion reduces the cached-input component for most Mini-SWE-agent configurations.
For Trae agent, cached-input cost decreases under reasoning deletion but increases under tool output deletion, where additional calls offset the reduction in retained content.
New-input cost stays close to the baseline, and re-billed input appears alongside it.
Under turn-level deletion, re-billed cost increases with deletion percentage because only selected turns invalidate the prefix.
Under token-level deletion, every call with an eligible turn invalidates the prefix.
At 25\%, the re-billed component of token-level deletion is about five times that of turn-level deletion.

The output component stays close to the uncompressed level for Mini-SWE-agent and grows for Trae agent, whose runs lengthen most.
Re-billed tokens contribute to every configuration's estimated cost increase, whereas reductions in cached-input cost provide limited or no offset.
Under tool output deletion, longer runs further increase token costs.
Delayed reasoning deletion produces no consistent increase in re-fetching or turns (Section~\ref{sec:rq2}), yet re-billing offsets the reduction in retained tokens.

\textbf{What this means for compression.}
Token counts decrease in 31 of the 42 delayed evaluations, and flat-pricing cost ratios fall below one in 26.
Our deletion configurations use no model, selector, or summary, yet token counts and flat pricing already suggest savings that the caching-aware model does not confirm.
Under the caching-aware model, none of the 42 delayed evaluations have lower estimated costs than the baseline, and 26 have significantly higher estimated costs.
Any approach that replaces an old turn breaks the cache at that turn as a deletion does, so we expect a compressor that rewrites before every call to pay the same re-billing, and one that rewrites once every several calls to pay it less often.

\begin{myframed}
\noindent
\textbf{Answer to RQ3.}
With the two-turn delay, deletion is significantly cheaper than the uncompressed runs in six of 42 evaluations of delayed configurations under flat pricing and in none under caching-aware pricing, where 26 evaluations are significantly more expensive.
The removed content otherwise qualifies for the cached-input price, while each rewrite incurs full input pricing from the first change onward.
Under delayed reasoning deletion, re-billing offsets the reduction in retained tokens.
Under tool output deletion, additional turns and completions further increase cost.
We find that token counts and flat pricing overstate what deletion saves, so a compression approach should be evaluated against the uncompressed agent with its prefix cache.
\end{myframed}

\noindent\textbf{Implications for future work.}
Our cost results call for studies on memory management that reduce total execution costs while preserving effectiveness under caching-aware pricing.
Evaluations should compare against an uncompressed baseline with the same cache assumptions and account for retrieval turns, generated reasoning, cache invalidation, and compression overhead.
Reporting tasks solved alongside cost, supplemented by cost per task solved, would clarify whether apparent savings accompany effectiveness losses.
Future work should also vary rewrite position and frequency to test whether less frequent rewriting or preserving longer prefixes lowers total cost after accounting for changes in agent behavior.

\section{Threats to Validity}
\label{sec:threats}

\noindent
A threat to the \textbf{internal validity} of our study is the variation between runs.
We use temperature zero, but LLM inference is not deterministic even then~\cite{ouyang2025:empirical}, and each configuration runs once per agent and task, with one random draw of what to delete.
The reference uses five uncompressed runs per task and agent, and the paired bootstrap resamples the 100 tasks.
The confidence intervals quantify task-sampling uncertainty only and exclude variation across repeated intervention runs or random deletion draws.
Each metric and timing also involves 42 paired comparisons, which raises the multiple-comparisons problem.
Our conclusions are based on patterns across the grid rather than on any single test, but as a sensitivity analysis, we also apply Holm’s correction within each family of 42 tests.
Fewer individual results remain significant, but the reported trends hold.
The results also depend on the experimental settings.
We use the agents' default prompts and tools, a 200-turn cap, and a two-turn delay taken from AgentDiet~\cite{xiao2026:reducing}. Different settings require additional evaluation.
Another threat to internal validity is data leakage~\cite{prathifkumar2025:does,liang2026:swebench}.
The repositories and their fixes are public, so the model may recall them, which would inflate the number of tasks solved.
At the same deletion percentage, reasoning deletion removes far fewer tokens than deletion of tool output or both scopes, confounding scope with the number of tokens removed.

A threat to \textbf{external validity} is the generalizability of our findings to other models, agents, tasks, and compression approaches.
We use one open-weight model, two agents with contrasting tool interfaces, and a 100-task subset of SWE-bench Verified.
The selected grid requires 9,400 runs and over 2,000 GPU hours, limiting the feasible benchmark size.
Exact prefix caching invalidates cached states after a modified token, but the resulting cost savings depend on provider pricing and cache availability.
The deletion configurations isolate the effects of random removal but do not represent learned selection or summarization approaches.
Learned selection might reduce information loss and re-fetching relative to random deletion. Infrequent rewriting also reduces the frequency of prefix invalidation. Neither is evaluated here.
An approach that rewrites an old turn still invalidates prefix reuse from the first change onward under the studied cost model.

Passing tests is a proxy for correctness but does not establish that a patch is the intended fix, threatening \textbf{construct validity}. We do not manually validate all 9,400 patches.
Another threat to construct validity is the cost model.
The caching-aware model omits cache-write premiums and minimum prefix lengths and assumes that every matching prefix is available. Expiry and eviction also limit this assumption. These simplifications can underestimate absolute costs. Their effects on the baseline and deletion configurations need not be equal, so the model does not establish provider-independent relative cost increases.
A last threat is the re-fetch measure.
The line-overlap measure misses known content returned in a different form, such as combined output from several source files.
We assess sensitivity to the 80\% threshold by repeating the analysis at 50\% and 90\%. The ratios differ, but the same configurations show the largest effects.

\section{Related Work}

\noindent
\textbf{Context management for agents.}
Approaches that reduce what an agent resends range from rules that mask old observations~\cite{yang2024:sweagent,lindenbauer2025:complexity}, through models that select what to remove~\cite{xiao2026:reducing,wang2026:swepruner,zeng2026:attncompress,franke2026:agentcontrolled}, to agents trained to manage their own context~\cite{liu2026contexttool,zhang2026:autocompact}.
More generally, long-horizon agents compress their observations and history~\cite{kang2026:acon,zhang2026:agora}, and prompt compression removes low-information tokens from a single prompt~\cite{jiang2023:llmlingua,pan2024:llmlingua2}, which resembles our token-level deletion except that it chooses tokens by their information content rather than at random.
These papers propose context-management approaches and evaluate tasks solved and tokens processed. Our study varies the scope, granularity, amount, and timing of random deletion to characterize the effects of these design choices and provide a baseline for context-management approaches.

\textbf{Trajectory reduction.}
Empirical work compares observation masking with summarization~\cite{lindenbauer2025:complexity} and prunes specific content such as reasoning~\cite{wang2026:when}.
Four recent studies are close to ours.
Satish et al.~\cite{satish2026:beyond} vary compression primitives from deployed agents across nearly~35,000 runs and report success, tokens, latency, and estimated cost.
Weinberger and Hozez~\cite{weinberger2026:token} measure provider-billed costs and find that removing 38\% of tool output tokens coincided with a 6.8\% higher cost.
Liu~\cite{liu2026:what} isolates the cost of reacquiring dropped states and shows that task completion can miss it. Franke~\cite{franke2026:agentcontrolled} lets agents evict tool results to a recoverable archive and notes that such edits can invalidate cached prefixes.
Our study complements these analyses by randomly varying deletion scope, granularity, amount, and timing, linking re-fetches to deleted versus retained outputs, and using complete prompt logs to reconstruct prefix invalidation and explain estimated cost changes.

\textbf{Behavior of agents.}
Recent work examines task success alongside common agent actions and recurring misbehavior patterns~\cite{bouzenia2025:understanding,liu2026:processcentric,ceka2026:understanding}, which can guide future agent runs~\cite{nanda2026:wink}.
Other work studies how users interact with agents and how that impacts the generated code~\cite{baumann2026:swechat}.
Our work, especially RQ2, inspects trajectories to understand how agents behave after trajectory reduction.

\section{Conclusion}
\label{sec:conclusion}

\noindent
This paper studies how trajectory deletion affects agent effectiveness, behavior, and estimated execution cost.
With a two-turn delay, 37 of 42 evaluations show no significant decrease in tasks solved. However, 21 evaluations have significantly longer runs, and the additional re-fetches primarily recover deleted tool output.
Immediate tool output deletion causes larger decreases in tasks solved than delayed deletion, particularly at token-level granularity. The inspected traces suggest that fragmented outputs sometimes induce attempts to repair apparently corrupted files.
Token reductions do not imply lower estimated execution costs.
Deletion yields estimated savings under flat pricing, but no delayed evaluation is cheaper under caching-aware pricing, and 26 of 42 across both agents are significantly more expensive than the uncompressed baseline.
Under the caching-aware model, deletion removes content that otherwise qualifies for discounted cached-input pricing, while each rewrite re-bills the suffix at the full input price.

The results have three practical implications.
The tested agents rely on recent tool outputs, supporting retention of these outputs until the agent uses them.
When recent tool output must be removed, turn-level deletion generally preserves more tasks solved than token-level deletion in the studied setting.
Compression approaches require comparison with an uncompressed baseline under caching-aware pricing, because token counts and flat pricing overstate the savings observed here.

The findings motivate evaluation of longer retention delays, selection based on observed reuse, and less frequent history rewriting. Batched rewriting still invalidates the prefix, so its net savings require evaluation and cannot be inferred from reduced rewrite frequency alone.

\medskip
\noindent
\textbf{Data Availability.}
The replication package~\cite{replicationpackage} contains all runs, every deleted turn's original text, and analysis code.

\printbibliography

@misc{anthropic2026contextediting,
  title = {Context Editing},
  author = {{Anthropic}},
  year = 2026,
  url = {https://platform.claude.com/docs/en/docs/build-with-claude/context-editing},
  langid = {english},
  howpublished = {\textsc{Url:} \url{https://platform.claude.com/docs/en/docs/build-with-claude/context-editing}}
}

@misc{anthropic2026promptcache,
  title = {Prompt Caching},
  author = {{Anthropic}},
  year = 2026,
  url = {https://platform.claude.com/docs/en/build-with-claude/prompt-caching},
  langid = {english},
  howpublished = {\textsc{Url:} \url{https://platform.claude.com/docs/en/build-with-claude/prompt-caching}}
}

@misc{anthropicpricing,
  title = {Pricing},
  author = {{Anthropic}},
  year = 2026,
  publisher = {Claude Platform Docs},
  url = {https://platform.claude.com/docs/en/about-claude/pricing},
  urldate = {2026-10-20},
  langid = {english},
  howpublished = {\textsc{Url:} \url{https://platform.claude.com/docs/en/about-claude/pricing}}
}

@misc{bai2026:how,
  title = {How Do {{AI}} Agents Spend Your Money? {{Analyzing}} and Predicting Token Consumption in Agentic Coding Tasks},
  shorttitle = {How Do {{AI}} Agents Spend Your Money?},
  author = {Bai, Longju and Huang, Zhemin and Wang, Xingyao and Sun, Jiao and Mihalcea, Rada and Brynjolfsson, Erik and Pentland, Alex and Pei, Jiaxin},
  year = 2026,
  month = oct,
  number = {arXiv:2604.22750},
  eprint = {2604.22750},
  primaryclass = {cs.CL},
  publisher = {arXiv},
  doi = {10.48550/arXiv.2604.22750},
  url = {http://arxiv.org/abs/2604.22750},
  urldate = {2026-10-23},
  archiveprefix = {arXiv},
  langid = {english}
}

@inproceedings{baumann2026:swechat,
  title = {{{SWE}}-Chat: Coding Agent Interactions from Real Users in the Wild},
  shorttitle = {{{SWE}}-Chat},
  booktitle = {Third {{Conference}} on {{Language Modeling}}},
  author = {Baumann, Joachim and Padmakumar, Vishakh and Li, Xiang and Yang, John and Yang, Diyi and Koyejo, Sanmi},
  options = {url=true},
  year = 2026,
  month = aug,
  url = {https://openreview.net/forum?id=KMAxjswrFe},
  urldate = {2026-10-23},
  langid = {english}
}

@inproceedings{bouzenia2025:understanding,
  title = {Understanding Software Engineering Agents: A Study of Thought-Action-Result Trajectories},
  shorttitle = {Understanding Software Engineering Agents},
  booktitle = {2025 40th {{IEEE}}/{{ACM International Conference}} on {{Automated Software Engineering}} ({{ASE}})},
  author = {Bouzenia, Islem and Pradel, Michael},
  year = 2025,
  month = nov,
  pages = {2846--2857},
  issn = {2643-1572},
  doi = {10.1109/ASE63991.2025.00234},
  url = {https://ieeexplore.ieee.org/document/11334353},
  urldate = {2026-10-23},
  langid = {english}
}

@article{ceka2026:understanding,
  title = {Understanding Automated Program Repair Agents through the Lens of Traceability: An Empirical Study},
  shorttitle = {Understanding Automated Program Repair Agents through the Lens of Traceability},
  author = {Ceka, Ira and Mitchell, Hailie and Pujar, Saurabh and Buratti, Luca and Ramji, Shyam and Yang, Junfeng and Kaiser, Gail and Ray, Baishakhi},
  year = 2026,
  month = oct,
  journal = {Proceedings of the ACM on Software Engineering},
  volume = {3},
  number = {ISSTA},
  pages = {ISSTA194:4450--ISSTA194:4473},
  doi = {10.1145/3832285},
  url = {https://dl.acm.org/doi/10.1145/3832285},
  urldate = {2026-10-23},
  langid = {english}
}

@misc{deepseek-ai2026:deepseekv4,
  title = {{{DeepSeek-V4}}: Towards Highly Efficient Million-Token Context Intelligence},
  shorttitle = {{{DeepSeek-V4}}},
  author = {{DeepSeek-AI} and Xu, Anyi and Lin, Bangcai and Xue, Bing and Wang, Bingxuan and Xu, Bingzheng and Wu, Bochao and Zhang, Bowei and Lin, Chaofan and Dong, Chen and Ling, Chenchen and Lu, Chengda and Zhao, Chenggang and Deng, Chengqi and Hou, Chengyu and Xu, Chenhao and Shao, Chenze and Ruan, Chong and Sun, Conner and Dai, Damai and Guo, Daya and Yang, Dejian and Chen, Deli and Li, Donghao and Ji, Dongjie and Li, Erhang and Wei, Fang and Lin, Fangyun and Yuan, Fangzhou and Xia, Feiyu and Dai, Fucong and Hao, Guangbo and Chen, Guanting and Cao, Guoai and Meng, Guolai and Li, Guowei and Yu, Han and Zhang, Han and Xu, Hanwei and Li, Hao and Liang, Haofen and Zhang, Haoling and Luo, Haoming and Wei, Haoran and Yuan, Haotian and Zhang, Haowei and Luo, Haowen and Chen, Haoyu and Ji, Haozhe and Zhang, Hengqing and Ding, Honghui and Tang, Hongxuan and Cao, Huanqi and Gao, Huazuo and Qu, Hui and Zeng, Hui and Yang, J. and Zhu, J. Q. and Luo, Jia and Song, Jia and Yu, Jia and Huang, Jialiang and Cai, Jialu and Liang, Jian and Zhou, Jiangting and Ye, Jiasheng and Li, Jiashi and Xu, Jiaxin and Hu, Jiewen and Yang, Jieyu and Chen, Jin and Yan, Jin and Chen, Jingchang and Zhou, Jingli and Xiang, Jingting and Yuan, Jingyang and Cheng, Jingyuan and Zhou, Jingzi and Zhu, Jinhua and Yu, Jiping and Sun, Joseph and Ran, Jun and Jiang, Junguang and Qiu, Junjie and Li, Junlong and Zheng, Junmin and Song, Junxiao and Dong, Kai and Gao, Kaige and Guan, Kang and Zhou, Kexing and Huang, Kezhao and Yu, Kuai and Wang, Lean and Zhang, Lecong and Wang, Lei and Xia, Leyi and Zhang, Li and Zhao, Liang and Guo, Lihua and Luo, Lingxiao and Ma, Linwang and Zhu, Linyan and Wang, Litong and Cai, Liyu and Zhang, Liyue and Chen, Longhao and Di, M. S. and Xu, M. Y. and Mei, Max and Wang, Miaojun and Zhang, Mingchuan and Zhang, Minghua and Tang, Minghui and Li, Mingming and Zhou, Mingxu and Han, Minmin and Wang, Ning and Huang, Panpan and Wang, Panpan and Cong, Peixin and Wang, Peiyi and Zhang, Peng and Wang, Qiancheng and Zhu, Qihao and Li, Qingyang and Chen, Qinyu and Du, Qiushi and Jiang, Qiwei and Tian, Rui and Xu, Ruifan and Lu, Ruijie and Xu, Ruiling and Ge, Ruiqi and Zhang, Ruisong and Pan, Ruizhe and Wang, Runji and Chen, Runqian and Yin, Runqiu and Xu, Runxin and Shen, Ruomeng and Zhang, Ruoyu and Chen, Ruyi and Liu, S. H. and Lu, Shanghao and Sun, Shangmian and Zhou, Shangyan and Chen, Shanhuang and Cai, Shaofei and Nie, Shaoheng and Wu, Shaoqing and Chen, Shaoyuan and Hu, Shengding and Liu, Shengyu and Hu, Shiqiang and Ma, Shirong and Wang, Shiyu and Yu, Shuiping and Zhou, Shunfeng and Pan, Shuting and Yu, Shuying and Zhou, Songyang and Ni, Tao and Yun, Tao and Jin, Tian and Pei, Tian and Ye, Tian and Lin, Tianle and Ji, Tianran and Cui, Tianyi and Yue, Tianyuan and Yu, Tingting and Wang, Tun and Zhang, W. and Xiao, W. L. and Zeng, Wangding and An, Wei and Zhao, Weilin and Liu, Wen and Liang, Wenfeng and Pang, Wenjie and Luo, Wenjing and Yao, Wenjing and Gao, Wenjun and Yang, Wenkai and Huang, Wenlve and Hou, Wenqing and Zhang, Wentao and Ma, Wenting and Gao, Xi and He, Xiang and Wang, Xiangwen and Wang, Xianzu and Bi, Xiao and Liu, Xiaodong and Wang, Xiaohan and Chen, Xiaokang and Zhang, Xiaokang and Nie, Xiaotao and Sun, Xiaowen and Wang, Xiaoxiang and Cheng, Xin and Liu, Xin and Xie, Xin and Liu, Xingchao and Liu, Xingchen and Yu, Xingkai and Li, Xingyou and Yang, Xinyu and Zhang, Xinyu and Chen, Xu and Wang, Xuanyu and Su, Xuecheng and Chen, Xueyin and Lin, Xuheng and Fu, Xuwei and Yan, Y. C. and Wang, Y. Q. and Ma, Y. W. and Luo, Yanfeng and Zhang, Yang and Xu, Yanhong and Ma, Yanru and Huang, Yanwen and Li, Yao and Li, Yao and Xu, Yao and Zhao, Yao and Sun, Yaofeng and Wang, Yaohui and Qian, Yi and Shao, Yi and Yu, Yi and Zhang, Yichao and Ding, Yifan and Shi, Yifan and Wu, Yijia and Xiong, Yiliang and Ma, Yiling and He, Ying and Tang, Ying and Zhou, Ying and Luo, Yingjia and Zhong, Yinmin and Piao, Yishi and Wang, Yisong and Zhang, Yixiang and Chen, Yixiao and Tan, Yixuan and Wei, Yixuan and Ma, Yiyang and Liu, Yiyuan and Yang, Yonglun and Guo, Yongqiang and Wu, Yongtong and Wu, Yu and Li, YuKun and Cheng, Yuan and Ou, Yuan and Xu, Yuanfan and Li, Yuanhao and Wang, Yuduan and Yang, Yuehan and Xu, Yuer and Wu, Yuhan and Meng, Yuhao and Zou, Yuheng and Zha, Yukun and Xiong, Yunfan and Chen, Yupeng and Lin, Yuping and Cao, Yuqian and Wang, Yuqian and Zhang, Yushun and Yan, Yuting and Lin, Yutong and Gu, Yuxian and Luo, Yuxiang and You, Yuxiang and Liu, Yuxuan and Zhou, Yuxuan and Zhou, Yuyang and Huang, Yuzhen and Wu, Z. F. and Wang, Zehao and Zhao, Zehua and Ren, Zehui and Zhang, Zekai and Sha, Zhangli and Fu, Zhe and Ju, Zhe and Xu, Zhean and Xie, Zhenda and Zhang, Zhengyan and Gao, Zheren and Hao, Zhewen and Gou, Zhibin and Ma, Zhicheng and Yan, Zhigang and Shao, Zhihong and Huang, Zhixian and Chen, Zhixuan and Wu, Zhiyu and Ren, Zhizhou and Wu, Zhongyu and Li, Zhuoshu and Zhang, Zhuping and Xu, Zian and Wang, Zihao and Qu, Zihua and Gu, Zihui and Zhu, Zijia and Li, Zilin and Zhang, Zipeng and Xie, Ziwei and Gao, Ziyi and Wan, Ziyi and Pan, Zizheng and Yao, Zongqing},
  year = 2026,
  month = apr,
  number = {arXiv:2606.19348},
  eprint = {2606.19348},
  primaryclass = {cs.CL},
  publisher = {arXiv},
  doi = {10.48550/arXiv.2606.19348},
  url = {http://arxiv.org/abs/2606.19348},
  urldate = {2026-10-20},
  archiveprefix = {arXiv},
  langid = {english}
}

@misc{deepseekpricing,
  title = {Models \& Pricing},
  author = {{DeepSeek}},
  year = 2026,
  publisher = {DeepSeek API Docs},
  url = {https://api-docs.deepseek.com/quick_start/pricing/},
  urldate = {2026-10-20},
  langid = {english},
  howpublished = {\textsc{Url:} \url{https://api-docs.deepseek.com/quick_start/pricing/}}
}

@misc{franke2026:agentcontrolled,
  title = {Agent-Controlled Forgetting for Tool-Using Agents: Reversible Context Curation in Practice},
  shorttitle = {Agent-Controlled Forgetting for Tool-Using Agents},
  author = {Franke, Jan-Peter},
  year = 2026,
  month = oct,
  number = {arXiv:2610.10590},
  eprint = {2610.10590},
  primaryclass = {cs.AI},
  publisher = {arXiv},
  doi = {10.48550/arXiv.2610.10590},
  url = {http://arxiv.org/abs/2610.10590},
  urldate = {2026-10-23},
  archiveprefix = {arXiv},
  langid = {english}
}

@misc{gao2025:trae,
  title = {Trae Agent: An {{LLM}}-Based Agent for Software Engineering with Test-Time Scaling},
  shorttitle = {Trae Agent},
  author = {Gao, Pengfei and Tian, Zhao and Meng, Xiangxin and Wang, Xinchen and Hu, Ruida and Xiao, Yuanan and Liu, Yizhou and Zhang, Zhao and Chen, Junjie and Gao, Cuiyun and Lin, Yun and Xiong, Yingfei and Peng, Chao and Liu, Xia},
  year = 2025,
  month = jul,
  number = {arXiv:2507.23370},
  eprint = {2507.23370},
  primaryclass = {cs},
  publisher = {arXiv},
  doi = {10.48550/arXiv.2507.23370},
  url = {http://arxiv.org/abs/2507.23370},
  urldate = {2026-05-27},
  archiveprefix = {arXiv},
  langid = {english}
}

@misc{googlepricing,
  title = {Gemini Developer {{API}} Pricing},
  author = {{Google}},
  year = 2026,
  publisher = {Google AI for Developers},
  url = {https://ai.google.dev/gemini-api/docs/pricing},
  urldate = {2026-10-20},
  langid = {english},
  howpublished = {\textsc{Url:} \url{https://ai.google.dev/gemini-api/docs/pricing}}
}

@inproceedings{jiang2023:llmlingua,
  title = {{{LLMLingua}}: Compressing Prompts for Accelerated Inference of Large Language Models},
  shorttitle = {{{LLMLingua}}},
  booktitle = {Proceedings of the 2023 {{Conference}} on {{Empirical Methods}} in {{Natural Language Processing}}},
  author = {Jiang, Huiqiang and Wu, Qianhui and Lin, Chin-Yew and Yang, Yuqing and Qiu, Lili},
  editor = {Bouamor, Houda and Pino, Juan and Bali, Kalika},
  year = 2023,
  month = dec,
  pages = {13358--13376},
  publisher = {Association for Computational Linguistics},
  address = {Singapore},
  doi = {10.18653/v1/2023.emnlp-main.825},
  url = {https://aclanthology.org/2023.emnlp-main.825/},
  urldate = {2026-10-23},
  langid = {english}
}

@inproceedings{jimenez2024swebench,
  title = {{{SWE}}-Bench: Can Language Models Resolve Real-World {{GitHub}} Issues?},
  shorttitle = {{{SWE}}-Bench},
  booktitle = {The {{Twelfth International Conference}} on {{Learning Representations}} ({{ICLR}})},
  author = {Jimenez, Carlos E. and Yang, John and Wettig, Alexander and Yao, Shunyu and Pei, Kexin and Press, Ofir and Narasimhan, Karthik R.},
  options = {url=true},
  year = 2024,
  url = {https://openreview.net/forum?id=VTF8yNQM66},
  langid = {english}
}

@inproceedings{kang2026:acon,
  title = {{{ACON}}: Optimizing Context Compression for Long-Horizon {{LLM}} Agents},
  shorttitle = {Acon},
  booktitle = {Forty-Third {{International Conference}} on {{Machine Learning}}},
  author = {Kang, Minki and Chen, Wei-Ning and Han, Dongge and Inan, Huseyin A. and Wutschitz, Lukas and Chen, Yanzhi and Sim, Robert and Rajmohan, Saravan},
  options = {url=true},
  year = 2026,
  month = jun,
  url = {https://openreview.net/forum?id=5EmOOLtH5P},
  langid = {english}
}

@inproceedings{kwon2023:efficient,
  title = {Efficient Memory Management for Large Language Model Serving with {{PagedAttention}}},
  booktitle = {Proceedings of the 29th {{Symposium}} on {{Operating Systems Principles}}},
  author = {Kwon, Woosuk and Li, Zhuohan and Zhuang, Siyuan and Sheng, Ying and Zheng, Lianmin and Yu, Cody Hao and Gonzalez, Joseph and Zhang, Hao and Stoica, Ion},
  year = 2023,
  month = oct,
  series = {{{SOSP}} '23},
  pages = {611--626},
  publisher = {Association for Computing Machinery},
  address = {New York, NY, USA},
  doi = {10.1145/3600006.3613165},
  url = {https://dl.acm.org/doi/10.1145/3600006.3613165},
  urldate = {2026-10-20},
  isbn = {979-8-4007-0229-7},
  langid = {english}
}

@inproceedings{liang2026:swebench,
  title = {The {{SWE}}-Bench Illusion: When State-of-the-Art {{LLMs}} Remember Instead of Reason},
  shorttitle = {The {{SWE}}-Bench Illusion},
  booktitle = {Proceedings of the {{IEEE}}/{{ACM}} 48th {{International Conference}} on {{Software Engineering}}: {{Software Engineering}} in {{Practice}}},
  author = {Liang, Shanchao and Garg, Spandan and Moghaddam, Roshanak Zilouchian},
  year = 2026,
  month = jul,
  series = {{{ICSE-SEIP}} '26},
  pages = {395--405},
  publisher = {Association for Computing Machinery},
  address = {New York, NY, USA},
  doi = {10.1145/3786583.3786882},
  url = {https://dl.acm.org/doi/10.1145/3786583.3786882},
  urldate = {2026-10-20},
  isbn = {979-8-4007-2426-8},
  langid = {english}
}

@inproceedings{lindenbauer2025:complexity,
  title = {The Complexity Trap: Simple Observation Masking Is as Efficient as {{LLM}} Summarization for Agent Context Management},
  shorttitle = {The Complexity Trap},
  booktitle = {{{NeurIPS}} 2025 {{Fourth Workshop}} on {{Deep Learning}} for {{Code}}},
  author = {Lindenbauer, Tobias and Slinko, Igor and Felder, Ludwig and Bogomolov, Egor and Zharov, Yaroslav},
  options = {url=true},
  year = 2025,
  month = nov,
  url = {https://openreview.net/forum?id=OHVzruJl5k},
  langid = {english}
}

@article{liu2026:processcentric,
  title = {Process-Centric Analysis of Agentic Software Systems},
  author = {Liu, Shuyang and Chen, Yang and Krishna, Rahul and Sinha, Saurabh and Ganhotra, Jatin and Jabbarvand, Reyhaneh},
  year = 2026,
  month = apr,
  journal = {Proceedings of the ACM on Programming Languages},
  volume = {10},
  number = {OOPSLA1},
  pages = {163:1961--163:1988},
  doi = {10.1145/3798271},
  url = {https://dl.acm.org/doi/10.1145/3798271},
  urldate = {2026-10-23},
  langid = {english}
}

@misc{liu2026:what,
  title = {What Does Context Compression Cost an Agent? {{Interaction}} Costs Unrevealed by Task-Completion Metrics},
  shorttitle = {What Does Context Compression Cost an Agent?},
  author = {Liu, Shuyu},
  year = 2026,
  month = aug,
  number = {arXiv:2608.16370},
  eprint = {2608.16370},
  primaryclass = {cs.AI},
  publisher = {arXiv},
  doi = {10.48550/arXiv.2608.16370},
  url = {http://arxiv.org/abs/2608.16370},
  urldate = {2026-10-23},
  archiveprefix = {arXiv},
  langid = {english}
}

@inproceedings{liu2026contexttool,
  title = {Context as a Tool: Context Management for Long-Horizon {{SWE}}-Agents},
  shorttitle = {Context as a Tool},
  booktitle = {Findings of the Association for Computational Linguistics: {{ACL}} 2026},
  author = {Liu, Shukai and Jiang, Bo and Yang, Jian and Li, Yizhi and Guo, Jinyang and Liu, Xianglong and Dai, Bryan},
  year = 2026,
  pages = {20604--20617},
  publisher = {Association for Computational Linguistics},
  doi = {10.18653/v1/2026.findings-acl.1032},
  url = {https://aclanthology.org/2026.findings-acl.1032/},
  langid = {english}
}

@inproceedings{nanda2026:wink,
  title = {Wink: Recovering from Misbehaviors in Coding Agents},
  shorttitle = {Wink},
  booktitle = {Proceedings of the 3rd {{ACM International Conference}} on {{AI-Powered Software}}},
  author = {Nanda, Rahul and Maddila, Chandra and Jha, Smriti and Khan, Euna Mehnaz and Paltenghi, Matteo and Chandra, Satish},
  year = 2026,
  month = jul,
  series = {{{AIware}} '26},
  pages = {208--217},
  publisher = {Association for Computing Machinery},
  address = {New York, NY, USA},
  doi = {10.1145/3805760.3814911},
  url = {https://dl.acm.org/doi/10.1145/3805760.3814911},
  urldate = {2026-10-23},
  isbn = {979-8-4007-2601-9},
  langid = {english}
}

@misc{openai2026promptcache,
  title = {Prompt Caching},
  author = {{OpenAI}},
  year = 2026,
  url = {https://developers.openai.com/api/docs/guides/prompt-caching},
  langid = {english},
  howpublished = {\textsc{Url:} \url{https://developers.openai.com/api/docs/guides/prompt-caching}}
}

@misc{openaipricing,
  title = {Pricing},
  author = {{OpenAI}},
  year = 2026,
  publisher = {OpenAI API Docs},
  url = {https://developers.openai.com/api/docs/pricing},
  urldate = {2026-10-20},
  langid = {english},
  howpublished = {\textsc{Url:} \url{https://developers.openai.com/api/docs/pricing}}
}

@article{ouyang2025:empirical,
  title = {An Empirical Study of the Non-Determinism of {{ChatGPT}} in Code Generation},
  author = {Ouyang, Shuyin and Zhang, Jie M. and Harman, Mark and Wang, Meng},
  year = 2025,
  journal = {ACM Transactions on Software Engineering and Methodology},
  volume = {34},
  number = {2},
  pages = {1--28},
  doi = {10.1145/3697010},
  langid = {english}
}

@inproceedings{pan2024:llmlingua2,
  title = {{{LLMLingua-2}}: Data Distillation for Efficient and Faithful Task-Agnostic Prompt Compression},
  shorttitle = {{{LLMLingua-2}}},
  booktitle = {Findings of the {{Association}} for {{Computational Linguistics}}: {{ACL}} 2024},
  author = {Pan, Zhuoshi and Wu, Qianhui and Jiang, Huiqiang and Xia, Menglin and Luo, Xufang and Zhang, Jue and Lin, Qingwei and R{\"u}hle, Victor and Yang, Yuqing and Lin, Chin-Yew and Zhao, H. Vicky and Qiu, Lili and Zhang, Dongmei},
  editor = {Ku, Lun-Wei and Martins, Andre and Srikumar, Vivek},
  year = 2024,
  month = aug,
  pages = {963--981},
  publisher = {Association for Computational Linguistics},
  address = {Bangkok, Thailand},
  doi = {10.18653/v1/2024.findings-acl.57},
  url = {https://aclanthology.org/2024.findings-acl.57/},
  urldate = {2026-10-23},
  langid = {english}
}

@misc{prathifkumar2025:does,
  title = {Does {{SWE}}-Bench-Verified Test Agent Ability or Model Memory?},
  author = {Prathifkumar, Thanosan and Mathews, Noble Saji and Nagappan, Meiyappan},
  year = 2025,
  month = dec,
  number = {arXiv:2512.10218},
  eprint = {2512.10218},
  primaryclass = {cs.SE},
  publisher = {arXiv},
  doi = {10.48550/arXiv.2512.10218},
  url = {http://arxiv.org/abs/2512.10218},
  urldate = {2026-10-20},
  archiveprefix = {arXiv},
  langid = {english}
}

@misc{replicationpackage,
  title = {Replication Package for ``The Hidden Costs of Forgetting: A Study of Trajectory Deletion in Software Engineering Agents''},
  author = {Vallecillos-Ruiz, Fernando and Bouzenia, Islem and Pradel, Michael and Moonen, Leon},
  year = 2026,
  publisher = {figshare},
  url = {https://figshare.com/s/72fce511c3ebe14d9dde},
  urldate = {2026-10-26},
  langid = {english},
  howpublished = {\textsc{Url:} \url{https://figshare.com/s/72fce511c3ebe14d9dde}}
}

@misc{satish2026:beyond,
  title = {Beyond Token Savings: A Systematic Study of Context Compression in {{LLM}} Agents},
  shorttitle = {Beyond Token Savings},
  author = {Satish, Ritul and Sinha, Prasoon and Kawada, Akiho and Yadwadkar, Neeraja J.},
  year = 2026,
  month = sep,
  number = {arXiv:2609.32961},
  eprint = {2609.32961},
  primaryclass = {cs.LG},
  publisher = {arXiv},
  doi = {10.48550/arXiv.2609.32961},
  url = {http://arxiv.org/abs/2609.32961},
  urldate = {2026-10-23},
  archiveprefix = {arXiv},
  langid = {english}
}

@misc{smith2025openhands,
  title = {{{OpenHands}} Context Condensensation for More Efficient {{AI}} Agents},
  author = {Smith, Calvin},
  year = 2025,
  url = {https://hub.openhands.dev/blog/openhands-context-condensensation-for-more-efficient-ai-agents},
  langid = {english},
  howpublished = {\textsc{Url:} \url{https://hub.openhands.dev/blog/openhands-context-condensensation-for-more-efficient-ai-agents}}
}

@misc{sweagent2026history,
  title = {History Processors},
  author = {{SWE-agent Team}},
  year = 2026,
  url = {https://swe-agent.com/latest/reference/history_processor_config/},
  langid = {english},
  howpublished = {\textsc{Url:} \url{https://swe-agent.com/latest/reference/history_processor_config/}}
}

@misc{vllm2026prefixcaching,
  title = {Automatic Prefix Caching},
  author = {{vLLM Project}},
  year = 2026,
  url = {https://docs.vllm.ai/en/latest/design/prefix_caching.html},
  langid = {english},
  howpublished = {\textsc{Url:} \url{https://docs.vllm.ai/en/latest/design/prefix_caching.html}}
}

@inproceedings{wang2025:openhands,
  title = {{{OpenHands}}: An Open Platform for {{AI}} Software Developers as Generalist Agents},
  shorttitle = {{{OpenHands}}},
  booktitle = {International {{Conference}} on {{Learning Representations}}},
  author = {Wang, Xingyao and Li, Boxuan and Song, Yufan and Xu, Frank F. and Tang, Xiangru and Zhuge, Mingchen and Pan, Jiayi and Song, Yueqi and Li, Bowen and Singh, Jaskirat and Tran, Hoang and Li, Fuqiang and Ma, Ren and Zheng, Mingzhang and Qian, Bill and Shao, Daniel and Muennighoff, Niklas and Zhang, Yizhe and Hui, Binyuan and Lin, Junyang and Brennan, Robert and Peng, Hao and Ji, Heng and Neubig, Graham},
  options = {url=true},
  year = 2025,
  month = may,
  volume = {2025},
  pages = {65882--65919},
  url = {https://proceedings.iclr.cc/paper_files/paper/2025/hash/a4b6ad6b48850c0c331d1259fc66a69c-Abstract-Conference.html},
  langid = {english}
}

@misc{wang2026:swepruner,
  title = {{{SWE}}-Pruner: Self-Adaptive Context Pruning for Coding Agents},
  shorttitle = {{{SWE}}-Pruner},
  author = {Wang, Yuhang and Shi, Yuling and Yang, Mo and Zhang, Rongrui and He, Shilin and Lian, Heng and Chen, Yuting and Ye, Siyu and Cai, Kai and Gu, Xiaodong},
  year = 2026,
  month = may,
  number = {arXiv:2601.16746},
  eprint = {2601.16746},
  primaryclass = {cs.SE},
  publisher = {arXiv},
  doi = {10.48550/arXiv.2601.16746},
  url = {http://arxiv.org/abs/2601.16746},
  urldate = {2026-10-23},
  archiveprefix = {arXiv},
  langid = {english}
}

@misc{wang2026:when,
  title = {When Can Agents Forget Their Reasoning? {{ICLR}} for Long-Horizon Agent Context Compression},
  shorttitle = {When Can Agents Forget Their Reasoning?},
  author = {Wang, Mingxuan and Luo, Fei and Wang, Bo and Yao, Guorun and Guo, Yinglong and Ning, Chao and Chen, Hongyue and Ma, Yanbiao and Han, Jungong},
  year = 2026,
  month = sep,
  number = {arXiv:2609.29875},
  eprint = {2609.29875},
  primaryclass = {cs.AI},
  publisher = {arXiv},
  doi = {10.48550/arXiv.2609.29875},
  url = {http://arxiv.org/abs/2609.29875},
  urldate = {2026-10-23},
  archiveprefix = {arXiv},
  langid = {english}
}

@misc{weinberger2026:token,
  title = {Token Reduction Is Not Cost Reduction},
  author = {Weinberger, Sarel and Hozez, Amir},
  year = 2026,
  month = aug,
  number = {arXiv:2607.12161},
  eprint = {2607.12161},
  primaryclass = {cs.CL},
  publisher = {arXiv},
  doi = {10.48550/arXiv.2607.12161},
  url = {http://arxiv.org/abs/2607.12161},
  urldate = {2026-10-20},
  archiveprefix = {arXiv},
  langid = {english}
}

@article{xiao2026:reducing,
  title = {Reducing Cost of {{LLM}} Agents with Trajectory Reduction},
  author = {Xiao, Yuan-An and Gao, Pengfei and Peng, Chao and Xiong, Yingfei},
  year = 2026,
  month = jun,
  journal = {Proceedings of the ACM on Software Engineering},
  volume = {3},
  number = {FSE},
  pages = {FSE056:1241--FSE056:1263},
  doi = {10.1145/3797084},
  url = {https://dl.acm.org/doi/10.1145/3797084},
  urldate = {2026-10-20},
  langid = {english}
}

@misc{xu2026:tokenpilot,
  title = {{{TokenPilot}}: Cache-Efficient Context Management for {{LLM}} Agents},
  shorttitle = {{{TokenPilot}}},
  author = {Xu, Buqiang and Xue, Zirui and Chen, Dianmou and Fu, Chenyang and Wu, Chiyu and Huang, Caiying and Jiang, Chen and Fang, Jizhan and Deng, Xinle and Chen, Yijun and Yao, Yunzhi and Wang, Xuehai and Shang, Jin and Yu, Gong and Zhang, Ningyu},
  year = 2026,
  month = aug,
  number = {arXiv:2606.17016},
  eprint = {2606.17016},
  primaryclass = {cs.CL},
  publisher = {arXiv},
  doi = {10.48550/arXiv.2606.17016},
  url = {http://arxiv.org/abs/2606.17016},
  urldate = {2026-10-23},
  archiveprefix = {arXiv},
  langid = {english}
}

@inproceedings{yang2024:sweagent,
  title = {{{SWE}}-Agent: Agent-Computer Interfaces Enable Automated Software Engineering},
  shorttitle = {{{SWE}}-Agent},
  booktitle = {Advances in {{Neural Information Processing Systems}} 37},
  author = {Yang, John and Jimenez, Carlos and Wettig, Alexander and Lieret, Kilian and Yao, Shunyu and Narasimhan, Karthik and Press, Ofir},
  year = 2024,
  pages = {50528--50652},
  publisher = {Neural Information Processing Systems Foundation, Inc. (NeurIPS)},
  address = {Vancouver, BC, Canada},
  doi = {10.52202/079017-1601},
  url = {http://www.proceedings.com/079017-1601.html},
  urldate = {2026-10-23},
  isbn = {979-8-3313-1438-5},
  langid = {english}
}

@article{zeng2026:attncompress,
  title = {{{AttnCompress}}: Dynamic Attention-Guided Trajectory Compression for Software Engineering Agents},
  shorttitle = {{{AttnCompress}}},
  author = {Zeng, Zhengran and Li, Yixin and Xie, Rui and Ye, Wei and Zhang, Shikun},
  year = 2026,
  month = oct,
  journal = {Proceedings of the ACM on Software Engineering},
  volume = {3},
  number = {ISSTA},
  pages = {ISSTA058:1331--ISSTA058:1353},
  doi = {10.1145/3832149},
  url = {https://dl.acm.org/doi/10.1145/3832149},
  urldate = {2026-10-23},
  langid = {english}
}

@misc{zhang2026:agora,
  title = {{{AGORA}}: Adapter-Grounded Observation-Action Retention for Inference-Free Prompt Compression in {{LLM}} Agents},
  shorttitle = {Agora},
  author = {Zhang, Haoran and Sun, Zhaohua},
  year = 2026,
  month = may,
  number = {arXiv:2605.26596},
  eprint = {2605.26596},
  primaryclass = {cs.AI},
  publisher = {arXiv},
  doi = {10.48550/arXiv.2605.26596},
  url = {http://arxiv.org/abs/2605.26596},
  urldate = {2026-10-23},
  archiveprefix = {arXiv},
  langid = {english}
}

@misc{zhang2026:autocompact,
  title = {{{AutoCompact}}: Learning When to Compact Context in Long-Horizon Coding Agents},
  shorttitle = {{{AutoCompact}}},
  author = {Zhang, Xuan and Zheng, Longtao and Du, Cunxiao and An, Bo and Dong, Xin},
  year = 2026,
  month = oct,
  number = {arXiv:2610.02163},
  eprint = {2610.02163},
  primaryclass = {cs.CL},
  publisher = {arXiv},
  doi = {10.48550/arXiv.2610.02163},
  url = {http://arxiv.org/abs/2610.02163},
  urldate = {2026-10-23},
  archiveprefix = {arXiv},
  langid = {english}
}

@inproceedings{zheng2024:sglang,
  title = {{{SGLang}}: Efficient Execution of Structured Language Model Programs},
  shorttitle = {{{SGLang}}},
  booktitle = {Advances in {{Neural Information Processing Systems}}},
  author = {Zheng, Lianmin and Yin, Liangsheng and Xie, Zhiqiang and Sun, Chuyue and Huang, Jeff and Yu, Cody Hao and Cao, Shiyi and Kozyrakis, Christos and Stoica, Ion and Gonzalez, Joseph E. and Barrett, Clark and Sheng, Ying},
  year = 2024,
  volume = {37},
  pages = {62557--62583},
  publisher = {Curran Associates, Inc.},
  doi = {10.52202/079017-2000},
  url = {https://proceedings.neurips.cc/paper_files/paper/2024/hash/724be4472168f31ba1c9ac630f15dec8-Abstract-Conference.html},
  urldate = {2026-10-23},
  langid = {english}
}

\end{document}